\documentclass[a4paper,11pt]{article}

\usepackage{amsmath,amsthm,amsbsy,amssymb,amsfonts,graphicx,mathrsfs,bm,accents,cite,xcolor,color}

\makeatletter
\catcode`\@=11
\@addtoreset{equation}{section}

\makeatother

\renewcommand{\thefootnote}{\arabic{footnote}}

\newcommand{\Exp}[1]{\operatorname{e}^{#1}}
\newcommand{\abs}[1]{\lvert{#1} \rvert}
\newcommand{\rmd}{{\mathrm{d}}}
\newcommand{\nn}{\nonumber}
\newcommand{\Lie}{\pounds}

\newcommand{\cB}{\mathcal B}

\newcommand{\cM}{\mathcal M}

\newcommand{\sfm}{\mathsf m}
\newcommand{\sfn}{\mathsf n}
\newcommand{\sfp}{\mathsf p}

\newcommand{\sfx}{\mathsf x}
\newcommand{\sfy}{\mathsf y}

\newcommand{\sfI}{\mathsf I}
\newcommand{\sfL}{\mathsf L}
\newcommand{\sfM}{\mathsf M}
\newcommand{\sfN}{\mathsf N}

\newcommand{\bbB}{\mathbb{B}}
\newcommand{\bbC}{\mathbb{C}}

\newcommand{\Ay}{{y}}
\newcommand{\Az}{{z}}
\newcommand{\By}{{\sfy}}
\newcommand{\rmT}{\mathrm{T}}
\newcommand{\GL}{\mathrm{GL}}
\newcommand{\SL}{\mathrm{SL}}
\newcommand{\Aphi}{\phi}
\newcommand{\Bphi}{\varphi}
\newcommand{\AG}{\mathfrak{g}}
\newcommand{\AB}{\mathcal{B}}
\newcommand{\AC}{\mathcal{C}}
\newcommand{\Aalpha}{\bm{\alpha}}
\newcommand{\Abeta}{\bm{\beta}}
\newcommand{\Agamma}{\bm{\gamma}}
\newcommand{\BG}{\mathsf{g}}
\newcommand{\BEG}{\mathsf{G}}
\newcommand{\BB}{\mathsf{B}}
\newcommand{\BC}{\mathsf{C}}
\newcommand{\Bc}{\mathsf{c}}
\newcommand{\BD}{\mathsf{D}}
\newcommand{\BR}{\mathsf{R}}
\newcommand{\BA}{\mathsf{A}}
\renewcommand{\AA}{\mathcal{A}}

\allowdisplaybreaks[3]

\begin{document}

\begin{titlepage}
\renewcommand{\thefootnote}{\fnsymbol{footnote}}

\vspace*{1.0cm}

\begin{center}
{\LARGE Connecting M-theory and type IIB parameterizations\\[3mm] in Exceptional Field Theory}%
\end{center}
\vspace{1.0cm}

\centerline{
{Yuho Sakatani$^{a,b}$}%
\footnote{E-mail address: \texttt{yuho@koto.kpu-m.ac.jp}}
\ \ and \ \ 
{Shozo Uehara$^{a}$}%
\footnote{E-mail address: \texttt{uehara@koto.kpu-m.ac.jp}}
}

\vspace{0.2cm}

\begin{center}
${}^a${\it Department of Physics, Kyoto Prefectural University of Medicine,}\\
{\it Kyoto 606-0823, Japan}\\
and\\
${}^b${\it Fields, Gravity \& Strings, CTPU}\\
{\it Institute for Basic Sciences, Daejeon 34047, Korea}
\end{center}
\vspace*{1cm}
\begin{abstract}
In the exceptional field theory, there are two natural parameterizations for the generalized metric; in terms of bosonic fields in the eleven-dimensional supergravity (M-theory parameterization) and the type IIB supergravity (type IIB parameterization). In order to translate various results known in the M-theory to the type IIB theory or vice versa, an explicit map between the two parameterizations will be useful. In this note, we present such a linear map. Comparing the two parameterizations under the linear map, we reproduce the known $T$-duality transformation rules for the supergravity fields. We also obtain the $T$-duality rules for non-geometric $\beta$-/$\gamma$-fields appearing in the type II theory on a torus. 
\end{abstract}
\thispagestyle{empty}
\end{titlepage}

\setcounter{footnote}{0}

\section{Introduction}

The Exceptional Field Theory (EFT) \cite{West:2001as,West:2003fc,West:2004st,Hillmann:2009ci,Berman:2010is,Berman:2011cg,Berman:2011jh,Berman:2012vc,Hohm:2013pua,Hohm:2013vpa,Hohm:2013uia,Aldazabal:2013via,Hohm:2014fxa} has been developed for a manifestly $U$-duality covariant formulation of supergravities. 
The generalized metric $\cM_{IJ}(x)$, which describes the geometry of the extended spacetime, is one of the most important fundamental fields in EFT. 
By providing a parameterization of $\cM_{IJ}(x)$ in terms of bosonic fields in the eleven-dimensional supergravity (M-theory parameterization) we can reproduce the action of the eleven-dimensional supergravity from the EFT action. 
On the other hand, if we consider a parameterization in terms of the bosonic fields in the type IIB supergravity (type IIB parameterization) the action of the type IIB supergravity can also be reproduced \cite{Hohm:2013pua,Blair:2013gqa}. 
In this sense, EFT successfully unifies the M-theory and the type IIB theory. 

The M-theory parameterization \cite{Berman:2011jh} takes the following structure:
\begin{align}
 \cM_{IJ} = (L_6^\rmT\,L_3^\rmT\,\hat{\cM}\,L_3\,L_6)_{IJ} \,,
\end{align}
where $\hat{\cM}$ is a block diagonal matrix consisting of the metric $G_{ij}$, while $L_3$ and $L_6$ are lower block-triangular matrices consisting of the 3-form potential $A_3$ and the 6-form potential $A_6$, respectively. 
On the other hand, the type IIB parameterization \cite{Tumanov:2014pfa,Lee:2016qwn} takes the form,
\begin{align}
 \sfM_{\sfM\sfN} = (\sfL_6^\rmT\,\sfL_4^\rmT\,\sfL_2^\rmT\, \hat{\sfM}\, \sfL_2\,\sfL_4\,\sfL_6)_{\sfM\sfN} \,.
\end{align}
Here, $\hat{\sfM}$ is a block diagonal matrix consisting of the Einstein-frame metric $\BEG_{\sfm\sfn}$ and scalar fields $(\Bphi,\,\BC_0)$, while $\sfL_2$, $\sfL_4$, and $\sfL_6$ are lower block-triangular matrices consisting of the 2-form potentials $(\BB_2,\,\BC_2)$, the 4-form potential $\BD_4$, and the 6-form potentials $(\bbB_6,\,\bbC_6)$, respectively.

In \cite{Sakatani:2016sko}, we proposed a worldvolume action for a $p$-brane, and the background fields are introduced only through the combination $\cM_{IJ}$. 
In order to compare our action with the actions for known branes, such as the M2-brane or the M5-brane, the explicit parameterizations for $\cM_{IJ}$ were important. 
Although only the M-theory branes are considered in \cite{Sakatani:2016sko}, according to the duality between the M-theory and the type IIB theory, we expect our brane action can also reproduce actions for the type IIB branes if we adopt the type IIB parameterization for the generalized metric. 
In order to realize this expectation, it is convenient to find a linear map between the M-theory parameterization and the type IIB parameterization. 

In this note, we find a constant linear map $S=(S^I{}_\sfM)$ that satisfies
\begin{align}
 \bigl(S^\rmT \cM\,S\bigr)_{\sfM\sfN} 
 = \cM_{IJ}\,S^I{}_\sfM\,S^J{}_\sfN = \sfM_{\sfM\sfN} 
\end{align}
with some identifications between the M-theory fields and the type IIB fields (see \cite{West:2004st,West:2004kb} and \cite{Malek:2015hma} for similar results connecting the type IIA and the type IIB theories). 
We show that, using the relation of fields in the eleven-dimensional supergravity and the type IIA supergravity, the above identifications are nothing but the $T$-duality transformation rules between the type IIA and the type IIB supergravity, known as Buscher's rules \cite{Buscher:1987sk,Buscher:1987qj,Bergshoeff:1995as,Meessen:1998qm,Eyras:1998hn}. 

In both the M-theory and the type IIB theory, we can also consider a different parameterization for $\cM_{IJ}$, called the non-geometric parameterization \cite{Andriot:2011uh,Andriot:2012wx,Andriot:2012an,Blumenhagen:2012nk,Blumenhagen:2012nt,Blumenhagen:2013aia,Andriot:2013xca,Andriot:2014qla,Lee:2016qwn}, in terms of the non-geometric potentials introduced in \cite{Duff:1989tf,Duff:1990hn}. 
Using the above linear map $S$, we find identifications of non-geometric potentials in the M-theory/type IIA supergravity and the type IIB supergravities. 
Namely, we obtain Buscher's rules for non-geometric potentials (see \cite{Lombardo:2016swq} for a recent work on a similar topic). 

\section{M-theory and type IIB parameterizations}

In this section, we fix our conventions for the M-theory parameterization and the type IIB parameterization (see also Appendix \ref{app:conventions} for our conventions). 

\subsection{M-theory parameterization}

Following \cite{West:2003fc,Berman:2011jh,Lee:2016qwn}, we consider the following decomposition of the generalized coordinates
\begin{align}
 (x^I) = (x^i\,,\ y_{i_1i_2}\,,\ y_{i_1\cdots i_5}\,,\ y_{i_1\cdots i_7,\,i}\,, \cdots) \qquad (i=1,\dotsc,d)\,,
\label{eq:M-theory-coordinates}
\end{align}
where $x^i$ are coordinates on the internal $d$-torus and the multiple indices are totally antisymmetrized and ellipses become relevant only for $E_{d(d)}$ with $d\geq 8$ (see section \ref{sec:summary} for comments on the $E_{8(8)}$ case). 
According to this decomposition, an explicit parameterization of the generalized metric for $d\leq 7$ is given by \cite{Berman:2011jh,Lee:2016qwn}
\begin{align}
 &\cM_{IJ} = (L_6^\rmT\,L_3^\rmT\,\hat{\cM}\,L_3\,L_6)_{IJ} \,,
\label{eq:M-theory-conventional}
\\
 &\hat{\cM} \equiv \abs{G}^{\frac{1}{n-2}}\,
 \begin{pmatrix}
 G_{ij} & 0 & 0 & 0 \\
 0 & G^{i_1i_2,\,j_1j_2} & 0 & 0 \\
 0 & 0 & G^{i_1\cdots i_5,\,j_1\cdots j_5} & 0 \\
 0 & 0 & 0 & G^{i_1\cdots i_7,\,j_1\cdots j_7}\,G^{ij}
 \end{pmatrix}\,, \quad 
 \abs{G}\equiv \det(G_{ij})\,,
\\
 &L_3 \equiv {\footnotesize
 \begin{pmatrix}
 \delta^i_j & 0 & 0 & 0 \\
 -\frac{A_{i_1i_2 j}}{\sqrt{2!}} & \delta_{i_1i_2}^{j_1j_2} & 0 & 0 \\
 -\frac{5!\,\delta_{i_1\cdots i_5}^{k_1\cdots k_5}\,A_{k_1k_2k_3}\, A_{k_4k_5 j}}{2!\,3!\,2!\sqrt{5!}} 
 & \frac{5!\, \delta^{j_1j_2k_1k_2k_3}_{i_1\cdots i_5}\, A_{k_1k_2k_3}}{3!\sqrt{2!\,5!}} 
 & \delta_{i_1\cdots i_5}^{j_1\cdots j_5} & 0 \\
 -\frac{7!\,\delta_{i_1\cdots i_7}^{k_1k_1p_1p_2p_3q_1q_2} \, A_{i k_1k_2}\,A_{p_1p_2p_3}\,A_{q_1q_2 j}}{3!\,2!\,3!\,2!\sqrt{7!}} 
 & \frac{7!\,\delta_{i_1\cdots i_7}^{j_1j_2 k_1k_2l_1l_2l_3}\, A_{i k_1k_2}\,A_{l_1l_2l_3}}{2!\,2!\,3!\sqrt{2!\,7!}} 
 & \frac{7!\,\delta^{j_1\cdots j_5k_1k_2}_{i_1\cdots i_7}\,A_{i k_1k_2}}{2!\sqrt{5!\,7!}} 
 & \delta_{i_1\cdots i_7}^{j_1\cdots j_7}\,\delta^j_i
 \end{pmatrix}}\,,
\\
 &L_6 \equiv {\footnotesize
 \begin{pmatrix}
 \delta^i_j & 0 & 0 & 0 
\\
 0 & \delta_{i_1i_2}^{j_1j_2} & 0 & 0 
\\
 \frac{A_{i_1\cdots i_5 j}}{\sqrt{5!}} & 0 & \delta_{i_1\cdots i_5}^{j_1\cdots j_5} & 0
\\
 0 & -\frac{7!\,\delta^{j_1j_2k_1\cdots k_5}_{i_1\cdots i_7}\,A_{i k_1\cdots k_5}}{5!\sqrt{2!\,7!}} & 0 & \delta_{i_1\cdots i_7}^{j_1\cdots j_7}\,\delta^j_i
 \end{pmatrix}}\,,
\end{align}
where $n\equiv 11-d$ is the dimension of the external space. 
Note that the sign of $y_{i_1\cdots i_5}$ is flipped compared to \cite{Lee:2016qwn}. 
Note also that $L_3$ and $L_6$ have the form,
\begin{align}
 &L_3 =\Exp{\ell_3}\,,\quad L_6 = \Exp{\ell_6}\,,
\\
 &\Exp{A}\equiv I+\sum_{n=1}^\infty\frac{1}{n!}\,A^n\,,\quad 
 I\equiv{\footnotesize
 \begin{pmatrix}
 \delta^i_j & 0 & 0 & 0 \\
 0 & \delta_{i_1i_2}^{j_1j_2} & 0 & 0 \\
 0 & 0 & \delta_{i_1\cdots i_5}^{j_1\cdots j_5} & 0\\
 0 & 0 & 0 & \delta_{i_1\cdots i_7}^{j_1\cdots j_7}\,\delta^j_i
 \end{pmatrix}}\,,
\\
 &\ell_3 \equiv {\footnotesize
 \begin{pmatrix}
 0 & 0 & 0 & 0 \\
 -\frac{3!\,\delta_{i_1i_2 j}^{l_1l_2l_3}}{\sqrt{2!}} & 0 & 0 & 0 \\
 0 & \frac{5!\, \delta^{j_1j_2 l_1l_2l_3}_{i_1\cdots i_5}}{\sqrt{2!\,5!}} & 0 & 0 \\
 0 & 0 & \frac{7!\,3\,\delta^{j_1\cdots j_5[l_1l_2}_{i_1\cdots i_7}\,\delta_i^{l_3]}}{\sqrt{5!\,7!}} & 0
 \end{pmatrix}\,\frac{A_{l_1l_2l_3}}{3!} }\,, 
\\
 &\ell_6 \equiv {\footnotesize
 \begin{pmatrix}
 0 & 0 & 0 & 0 \\
 0 & 0 & 0 & 0 \\
 \frac{6!\,\delta_{i_1\cdots i_5 j}^{l_1\cdots l_6}}{\sqrt{5!}} & 0 & 0 & 0\\
 0 & \frac{7!\,6\,\delta^{j_1j_2 [l_1\cdots l_5}_{i_1\cdots i_7}\,\delta_i^{l_6]} }{\sqrt{2!\,7!}} & 0 & 0
 \end{pmatrix}\,\frac{A_{l_1\cdots l_6}}{6!} }\,.
\end{align}
In our convention, components of a generalized vector are normalized as
\begin{align}
 (V^I)= \Bigl(v^i \quad \frac{v_{i_1i_2}}{\sqrt{2!}} \quad \frac{v_{i_1\cdots i_5}}{\sqrt{5!}} \quad \frac{v_{i_1\cdots i_7,\,i}}{\sqrt{7!}} \Bigr)^\rmT \,. 
\end{align}

\subsection{Type IIB parameterization}

In the type IIB case, we parameterize the generalized coordinates as \cite{Schnakenburg:2001he,West:2005gu,Tumanov:2014pfa,Lee:2016qwn}
\begin{align}
 (x^\sfM) = \bigl(\sfx^\sfm\,,\ \sfy_\sfm^\alpha\,,\ \sfy_{\sfm_1\sfm_2\sfm_3}\,,\ \sfy_{\sfm_1\cdots\sfm_5}^\alpha\,,\ \sfy_{\sfm_1\cdots \sfm_6,\,\sfm}\,, \cdots\bigr) \quad (\sfm=1,\dotsc,d-1\,,\ \alpha=1,\,2)\,,
\label{eq:IIB-coordinates}
\end{align}
where $\sfx^\sfm$ are coordinates on the internal $(d-1)$-torus and again the multiple indices are totally antisymmetrized and the ellipses become relevant only for $E_{d(d)}$ with $d\geq 8$\,. 

The type IIB parameterization of the generalized metric was found in \cite{Tumanov:2014pfa,Lee:2016qwn} from a group theoretical approach and the explicit form for $d\leq 7$ is given by
\begin{align}
 &\sfM_{\sfM\sfN} = (\sfL_6^\rmT\,\sfL_4^\rmT\,\sfL_2^\rmT\,\hat{\sfM}\,\sfL_2\,\sfL_4\,\sfL_6)_{\sfM\sfN} \,,\quad 
 \sfL_n \equiv \Exp{\ell_n}\quad (n=2,4,6)\,,
\label{eq:IIB-conventional}
\\
 &\Exp{A}\equiv \sfI+\sum_{n=1}^\infty\frac{1}{n!}\,A^n\,,\quad 
 \sfI\equiv {\arraycolsep=0.0mm {\footnotesize\left(\begin{array}{ccccc} 
 \delta^\sfm_\sfn & 0 & 0 & 0 & 0 \\
 0 & \delta^\alpha_\beta\,\delta_\sfm^\sfn & 0 & 0 & 0 \\
 0 & 0 & \delta_{\sfm_1\sfm_2\sfm_3}^{\sfn_1\sfn_2\sfn_3} & 0 & 0 \\
 0 & 0 & 0 & \delta^\alpha_\beta \,\delta_{\sfm_1\cdots \sfm_5}^{\sfn_1\cdots \sfn_5} & 0 \\
 0 & 0 & 0 & 0 & \delta_{\sfm_1\cdots\sfm_6}^{\sfn_1\cdots \sfn_6}\,\delta_\sfm^\sfn 
 \end{array}\right)}}\,,
\\
 &\hat{\sfM}\equiv \abs{\BEG}^{\frac{1}{n-2}}{\arraycolsep=0.5mm {\footnotesize\left(\begin{array}{ccccc} 
 \BEG_{\sfm\sfn} & 0 & 0 & 0 & 0 \\
 0 & m_{\alpha\beta} \,\BEG^{\sfm\sfn} & 0 & 0 & 0 \\
 0 & 0 & \BEG^{\sfm_1\sfm_2\sfm_3,\,\sfn_1\sfn_2\sfn_3} & 0 & 0 \\
 0 & 0 & 0 & m_{\alpha\beta} \,\BEG^{\sfm_1\cdots \sfm_5,\,\sfn_1\cdots \sfn_5} & 0 \\
 0 & 0 & 0 & 0 & \BEG^{\sfm_1\cdots\sfm_6,\,\sfn_1\cdots \sfn_6}\,\BEG^{\sfm\sfn}
 \end{array}\right)}} \,,
\\
 &\ell_2\equiv {\arraycolsep=0.5mm {\footnotesize\left(\begin{array}{ccccc} 
 0 & 0 & 0 & 0 & 0 \\
 2!\,\delta^\alpha_\gamma\,\delta_{\sfm\sfn}^{\sfp_1\sfp_2} & 0 & 0 & 0 & 0 \\
 0 & \frac{3!\,\epsilon_{\beta\gamma} \,\delta_{\sfm_1\sfm_2\sfm_3}^{\sfn \sfp_1\sfp_2}}{\sqrt{3!}} & 0 & 0 & 0 \\
 0 & 0 & \frac{5!\,\delta^\alpha_\gamma\,\delta_{\sfm_1\cdots \sfm_5}^{\sfn_1\sfn_2\sfn_3\sfp_1\sfp_2}}{\sqrt{3!\,5!}} & 0 & 0 \\
 0 & 0 & 0 & -\frac{6!\,2\,\epsilon_{\beta\gamma} \,\delta_{\sfm_1\cdots \sfm_6}^{\sfn_1\cdots \sfn_5[\sfp_1}\,\delta_{\sfm}^{\sfp_2]}}{\sqrt{5!\,6!}} & ~0~
 \end{array}\right)}\,\frac{\BB_{\sfp_1\sfp_2}^\gamma}{2!}} \,,
\\
 &\ell_4\equiv {\arraycolsep=0.5mm {\footnotesize\left(\begin{array}{ccccc} 
 0 & 0 & 0 & 0 & 0 \\
 0 & 0 & 0 & 0 & 0 \\
 \frac{4!\,\delta_{\sfm_1\sfm_2\sfm_3 \sfn}^{\sfp_1\cdots\sfp_4}}{\sqrt{3!}} & 0 & 0 & 0 & 0 \\
 0 & -\frac{5!\,\delta^\alpha_\beta \,\delta_{\sfm_1\cdots\sfm_5}^{\sfn\sfp_1\cdots\sfp_4}}{\sqrt{5!}} & 0 & 0 & 0 \\
 0 & 0 & -\frac{6!\,4\,\delta_{\sfm_1\cdots \sfm_6}^{\sfn_1\sfn_2\sfn_3[\sfp_1\sfp_2\sfp_3}\,\delta_{\sfm}^{\sfp_4]}}{\sqrt{3!\,6!}} & ~0~ & ~0~
 \end{array}\right)}\,\frac{\BD_{\sfp_1\cdots \sfp_4}}{4!}} \,, 
\\
 &\ell_6\equiv {\arraycolsep=0.5mm {\footnotesize\left(\begin{array}{ccccc} 
 0 & 0 & ~0~ & ~0~ & ~0~ \\
 0 & 0 & 0 & 0 & 0 \\
 0 & 0 & 0 & 0 & 0 \\
 \frac{6!\,\delta^\alpha_\gamma\,\delta_{\sfm_1\cdots \sfm_5 \sfn}^{\sfp_1\cdots\sfp_6}}{\sqrt{5!}} & 0 & 0 & 0 & 0 \\
 0 & \frac{6!\,6\,\epsilon_{\beta\gamma}\,\delta_{\sfm_1\cdots \sfm_6}^{\sfn[\sfp_1\cdots \sfp_5}\, \delta_{\sfm}^{\sfp_6]}}{\sqrt{6!}} & 0 & 0 & 0
 \end{array}\right)}\,\frac{\BD_{\sfp_1\cdots \sfp_6}^\gamma}{6!}} \,.
\end{align}
Note that the sign conventions for $\sfy_{\sfm_1\sfm_2\sfm_3}$, $\sfy_{\sfm_1\cdots \sfm_6,\,\sfm}$, and $\BB_{\sfm\sfn}^\alpha$ are opposite compared to \cite{Lee:2016qwn}. 
As we explain in section \ref{section:linear-map}, we rederive the above parameterization by using the M-theory parameterization \eqref{eq:M-theory-conventional} and making suitable identifications of the generalized coordinates \eqref{eq:M-theory-coordinates} and \eqref{eq:IIB-coordinates} together with suitable identifications of the background fields. 

Here, we shall explain each field appearing in $\sfM_{\sfM\sfN}$ (see \cite{Lee:2016qwn} for more details). 
The metric $\BEG_{\sfm\sfn}$ is (the internal part of) the Einstein-frame metric in the ten-dimensional type IIB supergravity. 
It is related to (the internal part of) the string-frame metric $\BG_{\sfm\sfn}$ by
\begin{align}
 \BEG_{\sfm\sfn} = \Exp{-\frac{1}{2}\,\Bphi}\BG_{\sfm\sfn}\,. 
\end{align}
Using the string-frame metric, $\hat{\sfM}$ becomes
\begin{align}
 \hat{\sfM} &= \abs{\BG}^{\frac{1}{n-2}} \Exp{-\frac{4}{n-2}\,\Bphi} 
\nn\\
 &\quad\times{\arraycolsep=0.5mm {\footnotesize\left(\begin{array}{ccccc} 
 \BG_{\sfm\sfn} & 0 & 0 & 0 & 0 \\
 0 & \Exp{\Bphi}m_{\alpha\beta} \,\BG^{\sfm\sfn} & 0 & 0 & 0 \\
 0 & 0 & \Exp{2\Bphi}\BG^{\sfm_1\sfm_2\sfm_3,\,\sfn_1\sfn_2\sfn_3} & 0 & 0 \\
 0 & 0 & 0 & \Exp{3\Bphi} m_{\alpha\beta} \,\BG^{\sfm_1\cdots \sfm_5,\,\sfn_1\cdots \sfn_5} & 0 \\
 0 & 0 & 0 & 0 & \Exp{4\Bphi}\,\BG^{\sfm_1\cdots \sfm_6,\,\sfn_1\cdots \sfn_6} \,\BG^{\sfm\sfn}
 \end{array}\right)}} .
\end{align}
The $\SL(2)$ matrix $m_{\alpha\beta}$ is given by
\begin{align}
 \bigl(m_{\alpha\beta}\bigr) &\equiv \Exp{\Bphi}\,\begin{pmatrix}
 \Exp{-2\Bphi} + (\BC_0)^2 & \BC_0 \\
 \BC_0 & 1
 \end{pmatrix}\,,\quad \det\bigl(m_{\alpha\beta}\bigr)=1 \,.
\end{align}
The antisymmetric symbols $\epsilon_{\alpha\beta}$ and $\epsilon^{\alpha\beta}$ are defined by $\epsilon_{12}=1=\epsilon^{12}$\,, and the inverse metric $m^{\alpha\beta}\equiv (m^{-1})^{\alpha\beta}$ satisfies $m^{\alpha\beta} = \epsilon^{\alpha\gamma}\,\epsilon^{\beta\delta}\,m_{\gamma\delta}$\,. 
A pair of 2-form potentials,
\begin{align}
 \BB^\alpha_{\sfm\sfn} \equiv \epsilon^{\alpha\beta}\, \BB_{\beta;\,\sfm\sfn}\,,\quad 
 (\BB_{\alpha;\,\sfm\sfn}) \equiv \begin{pmatrix} \BC_{\sfm\sfn}\\ \BB_{\sfm\sfn} \end{pmatrix} \,,
\end{align}
is an $\SL(2)$ $S$-duality doublet and the 4-form potential $\BD_4$ is $S$-duality invariant. 
We can define another Ramond-Ramond 4-form $\BC_4$ as $\BC_4 \equiv \BD_4 + \frac{1}{2}\,\BB_2\wedge\BC_2$, or, in components,
\begin{align}
 \BC_{\sfm_1\cdots \sfm_4}\equiv \BD_{\sfm_1\cdots \sfm_4} + 3\,\BB_{[\sfm_1\sfm_2}\, \BC_{\sfm_3\sfm_4]} \,.
\end{align}
A pair of 6-form potentials,
\begin{align}
 \BD^\alpha_{\sfm_1\cdots \sfm_6} \equiv \epsilon^{\alpha\beta}\,\BD_{\beta;\,\sfm_1\cdots \sfm_6}\,,\quad 
 (\BD_{\alpha;\,\sfm_1\cdots \sfm_6}) \equiv \begin{pmatrix} \bbB_{\sfm_1\cdots \sfm_6}\\ \bbC_{\sfm_1\cdots \sfm_6} \end{pmatrix} \,,
\end{align}
transforms as an $S$-duality doublet. 
We also define $(\BC_{\sfm_1\cdots \sfm_6},\,\BB_{\sfm_1\cdots \sfm_6})$ as \cite{Lee:2016qwn}\footnote{Note that the sign conventions of $\BB_{\sfm\sfn}$, $\bbB_{\sfm_1\cdots \sfm_6}$, and $\BB_{\sfm_1\cdots \sfm_6}$ are also different from \cite{Lee:2016qwn}.}
\begin{align}
\begin{split}
 \BC_{\sfm_1\cdots \sfm_6}
 &\equiv \bbC_{\sfm_1\cdots \sfm_6}+15\,\BD_{[\sfm_1\cdots \sfm_4}\,\BB_{\sfm_5\sfm_6]} + 15\, \BB_{[\sfm_1\sfm_2}\,\BB_{\sfm_3\sfm_4}\, \BC_{\sfm_5\sfm_6]}\,,
\\
 \BB_{\sfm_1\cdots \sfm_6}
 &\equiv \bbB_{\sfm_1\cdots \sfm_6} + 15\,\BD_{[\sfm_1\cdots \sfm_4}\,\BC_{\sfm_5\sfm_6]} +30\,\BC_{[\sfm_1\sfm_2}\,\BC_{\sfm_3\sfm_4}\,\BB_{\sfm_5\sfm_6]}\,.
\end{split}
\end{align}
The $\SL(2)$ transformation rules are summarized as
\begin{align}
\begin{split}
 \BEG'_{\sfm\sfn}&=\BEG_{\sfm\sfn}\,,\quad 
 m'_{\alpha\beta}= (\Lambda^\rmT)_\alpha{}^\gamma\,m_{\gamma\delta}\, \Lambda^\delta{}_\alpha \quad \Bigl(\tau'=\frac{a\,\tau+b}{c\,\tau+d} \,,\quad \tau\equiv \BC_0+i \Exp{-\Bphi}\Bigr)\,,
\\
 \BB'^{\alpha}_2 &= (\Lambda^{-1})^\alpha{}_\beta\,\BB_2^\beta \,,\quad 
 \BD'_4=\BD_4\,,\quad 
 \BD'^{\alpha}_6 = (\Lambda^{-1})^\alpha{}_\beta\,\BD_6^\beta \,,\quad 
 \Lambda\equiv\Bigl(\begin{smallmatrix} a & c \\ b & d \end{smallmatrix}\Bigr)\quad (a\,d-b\,c=1)\,.
\end{split}
\label{eq:SL2-IIB}
\end{align}
In particular, the $S$-duality transformation corresponds to $\Lambda = \bigl(\begin{smallmatrix} 0 & -1 \\ 1 & 0 \end{smallmatrix}\bigr)$.
From these we obtain the $S$-duality transformation rules,
\begin{align}
\begin{split}
 &\BB'_2=-\BC_2\,,\quad \BC'_2=\BB_2\,,\quad \BC'_4 = \BC_4 - \BB_2\wedge \BC_2 \,, 
\\
 &\BC'_6 = - \Bigl(\BB_6 - \frac{1}{2}\,\BB_2\wedge \BC_2\wedge \BC_2\Bigr)\,,\quad 
 \BB'_6 = \BC_6 - \frac{1}{2}\,\BC_2\wedge \BB_2\wedge \BB_2\,. 
\end{split}
\end{align}
A generalized vector in the type IIB parameterization is normalized as
\begin{align}
 (V^\sfM)= \Bigl(v^\sfm \quad v_\sfm^\alpha \quad
 \frac{v_{\sfm_1\sfm_2\sfm_3}}{\sqrt{3!}} \quad
 \frac{v_{\sfm_1\cdots \sfm_5}^\alpha}{\sqrt{5!}} \quad
 \frac{v_{\sfm_1\cdots \sfm_6,\,\sfm}}{\sqrt{6!}} \Bigr)^\rmT \,. 
\end{align}

\section{Linear map between M-theory/type IIB parameterizations}
\label{section:linear-map}

In this section, we construct an explicit linear map between the two parameterizations. 
In order to find a map, we decompose the physical coordinates on the M-theory side as
\begin{align}
 (x^i)= (x^a,\,x^\alpha)\qquad (a=1,\dotsc,d-2\,,\ \alpha =\Ay,\,\Az) \,,
\end{align}
where $x^\Az$ is supposed to be the coordinate on the M-theory circle and $x^\Ay$ is an arbitrary coordinate among the remaining $(d-1)$ coordinates. 
On the other hand, on the type IIB side, we consider the following $(d-2)+1$ decomposition:
\begin{align}
 (\sfx^\sfm)= (\sfx^a,\,\sfx^\By)\qquad (a=1,\dotsc,d-2) \,.
\end{align}
We then find a map between the M-theory compactified on a 2-torus (with coordinates $x^\alpha$) and the type IIB theory compactified on a circle (with a coordinate $\sfx^\By$). 
This map precisely corresponds to the duality proposed in \cite{Schwarz:1995dk,Schwarz:1995jq} and the $\SL(2)$ symmetry in the type IIB theory \eqref{eq:SL2-IIB} corresponds to the $\SL(2)$ symmetry in the M-theory which changes the modular parameter on the 2-torus. 

\subsection{Linear map}

We consider the following $\GL(d-2)\times \SL(2)$-covariant linear map between generalized vectors in the two parameterizations (see Appendix \ref{app:outline} for the convention for the combinatoric factors):
\begin{align}
 \begin{pmatrix}
 v^a \\[-2mm] v^{\By} \\ \hline v^\alpha_a \\[-2mm] v^\alpha_{\By} \\ \hline \frac{1}{\sqrt{3!}}\,v_{a_1a_2a_3} \\[-2mm] \frac{1}{\sqrt{2!}}\,v_{a_1a_2 \By} \\ \hline \frac{1}{\sqrt{5!}}\,v^\alpha_{a_1\cdots a_5} \\[-2mm] \frac{1}{\sqrt{4!}}\,v^\alpha_{a_1\cdots a_4\By} \\ \hline \frac{1}{\sqrt{5!}}\,v_{a_1\cdots a_5\By,a} \\[-2mm] \frac{1}{\sqrt{5!}}\,v_{a_1\cdots a_5\By,\,\By}
 \end{pmatrix}_{\!\!\!\text{(IIB)}}
 \hspace{-15pt}
 = (S^{-1})^\sfM{}_J
 \begin{pmatrix}
 v^b \\[-2mm] v^\beta \\ \hline \frac{1}{\sqrt{2!}}\,v_{b_1b_2} \\[-2mm] v_{b \beta} \\[-2mm] v_{\Ay\Az} \\ \hline \frac{1}{\sqrt{5!}}\,v_{b_1\cdots b_5} \\[-2mm] \frac{1}{\sqrt{4!}}\,v_{b_1\cdots b_4\beta} \\[-2mm] \frac{1}{\sqrt{3!}}\,v_{b_1b_2b_3\Ay\Az} \\ \hline \frac{1}{\sqrt{5!}}\,v_{b_1\cdots b_5\Ay\Az,\,b} \\[-2mm] \frac{1}{\sqrt{5!}}\,v_{b_1\cdots b_5\Ay\Az,\,\beta}
 \end{pmatrix}_{\!\!\!\text{(M)}}\,,
\end{align}
where
\begin{align}
 (S^{-1})^\sfM{}_J &\equiv {\footnotesize {\arraycolsep=0.7mm \left(\begin{array}{cc|ccc|ccc|cc}
 \delta^a_b & 0 & 0 & 0 & 0 & 0 & 0 & 0 & 0 & 0 \\
 0 & 0 & 0 & 0 & ~1~ & 0 & 0 & 0 & 0 & 0 \\ \hline
 0 & 0 & 0 & \epsilon^{\alpha\beta}\,\delta_a^b & 0 & 0 & 0 & 0 & 0 & 0 \\
 0 & \delta^\alpha_\beta & 0 & 0 & 0 & 0 & 0 & 0 & 0 & 0 \\ \hline
 0 & 0 & 0 & 0 & 0 & 0 & 0 & \delta_{a_1a_2a_3}^{b_1b_2b_3} & 0 & 0 \\
 0 & 0 & \delta_{a_1a_2}^{b_1b_2} & 0 & 0 & 0 & 0 & 0 & 0 & 0 \\ \hline
 0 & 0 & 0 & 0 & 0 & 0 & 0 & 0 & 0 & \epsilon^{\alpha\beta}\,\delta^{b_1\cdots b_5}_{a_1\cdots a_5} \\
 0 & 0 & 0 & 0 & 0 & 0 & \epsilon^{\alpha\beta}\,\delta_{a_1\cdots a_4}^{b_1\cdots b_4} & 0 & 0 & 0 \\ \hline
 0 & 0 & 0 & 0 & 0 & 0 & 0 & 0 & \delta_{a_1\cdots a_5}^{b_1\cdots b_5}\,\delta_a^b & 0 \\
 0 & 0 & 0 & 0 & 0 & \delta_{a_1\cdots a_5}^{b_1\cdots b_5} & 0 & 0 & 0 & 0 
 \end{array}\right)}} \,,
\\
 S^I{}_\sfN &\equiv {\footnotesize {\arraycolsep=0.7mm \left(\begin{array}{cc|cc|cc|cc|cc}
 \delta^a_b & 0 & 0 & 0 & 0 & 0 & 0 & 0 & 0 & 0 \\
 0 & 0 & 0 & \delta^\alpha_\beta & 0 & 0 & 0 & 0 & 0 & 0 \\ \hline
 0 & 0 & 0 & 0 & 0 & \delta_{a_1a_2}^{b_1b_2} & 0 & 0 & 0 & 0 \\
 0 & 0 & \epsilon^\rmT_{\alpha\beta}\,\delta_a^b & 0 & 0 & 0 & 0 & 0 & 0 & 0 \\
 0 & ~1~ & 0 & 0 & 0 & 0 & 0 & 0 & 0 & 0 \\ \hline
 0 & 0 & 0 & 0 & 0 & 0 & 0 & 0 & 0 & \delta_{a_1\cdots a_5}^{b_1\cdots b_5} \\
 0 & 0 & 0 & 0 & 0 & 0 & 0 & \epsilon^\rmT_{\alpha\beta}\,\delta_{a_1\cdots a_4}^{b_1\cdots b_4} & 0 & 0 \\
 0 & 0 & 0 & 0 & \delta_{a_1a_2a_3}^{b_1b_2b_3} & 0 & 0 & 0 & 0 & 0 \\ \hline
 0 & 0 & 0 & 0 & 0 & 0 & 0 & 0 & \delta_{a_1\cdots a_5}^{b_1\cdots b_5}\,\delta_a^b & 0 \\
 0 & 0 & 0 & 0 & 0 & 0 & \epsilon^\rmT_{\alpha\beta} \,\delta_{a_1\cdots a_5}^{b_1\cdots b_5} & 0 & 0 & 0 
 \end{array}\right)}} \,.
\end{align}
In terms of the generalized coordinates, we consider the following mappings:
\begin{align}
 \begin{pmatrix}
 \sfx^a \\[-2mm] \sfx^{\By} \\ \hline 
 \sfy^\alpha_a \\[-2mm] \sfy^\alpha_{\By} \\ \hline
 \sfy_{a_1a_2a_3} \\[-2mm] \sfy_{a_1a_2 \By} \\ \hline \sfy^\alpha_{a_1\cdots a_5} \\[-2mm] \sfy^\alpha_{a_1\cdots a_4\By} \\ \hline 
 \sfy_{a_1\cdots a_5\By,\,b} \\[-2mm]
 \sfy_{a_1\cdots a_5\By,\,\By}
 \end{pmatrix}_{\!\!\!\text{(IIB)}} \hspace{-10pt}
 = \begin{pmatrix}
 x^a \\[-2mm]
 y_{\Ay\Az} \\ \hline
 \epsilon^{\alpha\beta}\,y_{a\beta} \\[-2mm]
 x^\alpha \\ \hline
 y_{a_1a_2a_3\Ay\Az} \\[-2mm] 
 y_{a_1a_2} \\ \hline
 \epsilon^{\alpha\beta}\,y_{a_1\cdots a_5 \Ay\Az,\,\beta} \\[-2mm] 
 \epsilon^{\alpha\beta}\,y_{a_1\cdots a_4\beta} \\ \hline
 y_{a_1\cdots a_5\Ay\Az,\,b} \\[-2mm]
 y_{a_1\cdots a_5}
 \end{pmatrix}_{{\!\!\!\text{(M)}}} \quad \text{or}\quad 
 \begin{pmatrix}
 x^a \\[-2mm] x^\alpha \\ \hline y_{a_1a_2} \\[-2mm] y_{a \alpha} \\[-2mm] y_{\Ay\Az} \\ \hline y_{a_1\cdots a_5} \\[-2mm] y_{a_1\cdots a_4\alpha} \\[-2mm] y_{a_1a_2a_3 \Ay\Az} \\ \hline 
 y_{a_1\cdots a_5\Ay\Az,\,b} \\[-2mm] 
 y_{a_1\cdots a_5 \Ay\Az,\,\alpha}
 \end{pmatrix}_{\!\!\!\text{(M)}}
 \hspace{-10pt}
 =
\begin{pmatrix}
 \sfx^a \\[-2mm]
 \sfy_\By^\alpha \\ \hline
 \sfy_{a_1a_2\By} \\[-2mm]
 \epsilon^\rmT_{\alpha\beta}\,\sfy_a^\beta \\[-2mm]
 \sfx^\By \\ \hline
 \sfy_{a_1\cdots a_5\By,\,\By} \\[-2mm]
 \epsilon^\rmT_{\alpha\beta}\,\sfy^\beta_{a_1\cdots a_4\By} \\[-2mm]
 \sfy_{a_1a_2a_3} \\ \hline
 \sfy_{a_1\cdots a_5\By,\,b} \\[-2mm]
 \epsilon^\rmT_{\alpha\beta}\,\sfy_{a_1\cdots a_5}^\beta 
\end{pmatrix}_{{\!\!\!\text{(IIB)}}} \,.
\label{eq:map-coordinates}
\end{align}
Note that some of the above identifications appear in (6.10) of \cite{West:2004st} and (4.7) and (4.8) of \cite{West:2004kb}, where the identifications are found from a different argument. 

After some tedious calculation (see Appendix \ref{app:outline} for the details), one can show that the generalized metric $\cM_{IJ}$ given in \eqref{eq:M-theory-conventional} is mapped to $\sfM_{\sfM\sfN}$ given in \eqref{eq:IIB-conventional}, namely,
\begin{align}
 \bigl(S^\rmT\cM\,S\bigr)_{\sfM\sfN} = \sfM_{\sfM\sfN} \,,
\label{eq:linear-map}
\end{align}
under the following identifications:
\begin{align}
\begin{split}
 &\bigl(G_{ij}\bigr) 
 = \begin{pmatrix}
 G_{ab} & G_{a\beta} \\ G_{\alpha b} & G_{\alpha\beta}
 \end{pmatrix}
 = \Exp{-\frac{2}{3}\,\Bphi}\,\BG_{\By\By}^{1/3}
 \begin{pmatrix}
 \delta_a^c & -\BB_{a\By}^\gamma \\ 0 & \delta_\alpha^\gamma
 \end{pmatrix}
\begin{pmatrix}
 \frac{2\,\BG_{c\By,\,d\By}}{\BG_{\By\By}} & 0 \\ 0 & \frac{\Exp{\Bphi}}{\BG_{\By\By}}\, m_{\gamma\delta}
 \end{pmatrix}
 \begin{pmatrix}
 \delta^d_b & 0 \\ -\BB_{b\By}^\delta & \delta^\delta_\beta
 \end{pmatrix} \,,
\\
 &A_{abc} = \widetilde{\BD}_{abc \By} \,, \qquad
 A_{ab \alpha} = \epsilon^\rmT_{\alpha\beta} \, \widetilde{\BB}^\beta_{ab} \,, \qquad 
 A_{a \Ay\Az} = \frac{\BG_{a \By}}{\BG_{\By\By}} \,,
\\
 &A_{a_1\cdots a_5 \alpha} = \epsilon^\rmT_{\alpha\beta}\,\widetilde{\BD}^\beta_{a_1\cdots a_5 \By} \,,\qquad 
 A_{a_1\cdots a_4 \Ay\Az} = \widetilde{\BD}_{a_1\cdots a_4} \,,
\label{eq:M-IIB-map}
\end{split}
\end{align}
where we defined
\begin{align}
\begin{split}
 \widetilde{\BD}_{abc \By} &\equiv \BD_{abc \By} -\frac{3}{2}\,\epsilon_{\gamma\delta}\,\BB^\gamma_{[ab}\,\BB^\delta_{c]\By}
 - \frac{3\,\epsilon_{\gamma\delta}\,\BB^\gamma_{[a|\By|}\,\BB^\delta_{b|\By|}\,\BG_{c]\By}}{\BG_{\By\By}} \,,
\\
 \widetilde{\BB}^\alpha_{a_1a_2} &\equiv \BB^\alpha_{a_1a_2} - \frac{\BB^\alpha_{a_1 \By}\,\BG_{a_2 \By} - \BG_{a_1 \By}\,\BB^\alpha_{a_2 \By}}{\BG_{\By\By}} \,,
\\
 \widetilde{\BD}^\alpha_{a_1\cdots a_5 \By}&\equiv \BD^\alpha_{a_1\cdots a_5 \By} + 5\,\BD_{[a_1a_2a_3|\By|}\,\BB^\alpha_{a_4a_5]} + \frac{5}{2}\, \epsilon_{\gamma\delta}\,\BB^\alpha_{[a_1a_2}\,\BB^\gamma_{a_3a_4}\,\BB^\delta_{a_5]\By}
\\
 &\quad - \frac{10\,\BD_{[a_1\cdots a_3|\By|}\,\BB^\alpha_{a_4|\By|}\BG_{a_5]\By}}{\BG_{\By\By}} + \frac{15\, \epsilon_{\gamma\delta}\,\BB^\alpha_{[a_1a_2}\, \BB^\gamma_{a_3|\By|}\,\BB^\delta_{a_4|\By|}\,\BG_{a_5]\By}}{2\,\BG_{\By\By}} \,,
\\
 \widetilde{\BD}_{a_1\cdots a_4}&\equiv \BD_{a_1\cdots a_4} - \frac{2\,\BD_{[a_1a_2a_3|\By|}\,\BG_{a_4]\By}}{\BG_{\By\By}} -\frac{3\,\epsilon_{\gamma\delta}\,\BB^\gamma_{[a_1a_2}\,\BB^\delta_{a_3|\By|} \, \BG_{a_4]\By}}{\BG_{\By\By}} \,.
\end{split}
\end{align}

As we show below, the identifications \eqref{eq:M-IIB-map} are precisely Buscher's rules after rewriting the fields in the eleven-dimensional supergravity in terms of those in the type IIA supergravity.

\subsection{Conventional Buscher's rules}

The supergravity fields in the eleven-dimensional and the type IIA supergravity are related by
\begin{align}
\begin{split}
 (G_{ij})&\equiv \begin{pmatrix}
 G_{rs} & G_{r \Az} \\ G_{\Az s} & G_{\Az\Az}
 \end{pmatrix}
 =\begin{pmatrix}
 \Exp{-\frac{2}{3}\,\Aphi}\,\AG_{rs}+\Exp{\frac{4}{3}\,\Aphi}\,\AC_r\, \AC_s & \Exp{\frac{4}{3}\,\Aphi}\, \AC_r \\ \Exp{\frac{4}{3}\,\Aphi}\, \AC_s & \Exp{\frac{4}{3}\,\Aphi}
 \end{pmatrix} \,,
\\
 A_{pqr} &= \AC_{pqr} \,,\qquad 
 A_{pq \Az} = \AB_{pq} \,,
\\
 A_{p_1\cdots p_6} &= \AB_{p_1\cdots p_6} \,,\qquad 
 A_{p_1\cdots p_5\Az} = \AC_{p_1\cdots p_5} + 5\,\AC_{[p_1p_2p_3}\,\AB_{p_4p_5]} \,, 
\label{eq:IIA-reduction}
\end{split}
\end{align}
where $p,q,r=0,\dotsc,9$, and recall that $x^\Az$ is the coordinate on the M-theory circle. 
Plugging these into \eqref{eq:M-IIB-map}, we obtain the following relations: 
\begin{align}
 \AG_{ab} &= \BG_{ab} - \frac{\BG_{a \By}\,\BG_{b \By}-\BB_{a \By}\,\BB_{b \By}}{\BG_{\By\By}}\,,\qquad 
 \AG_{a \Ay}=-\frac{\BB_{a \By}}{\BG_{\By\By}}\,,\qquad 
 \AG_{\Ay\Ay}=\frac{1}{\BG_{\By\By}}\,,
\\
 \AB_{ab} &= \BB_{ab} - \frac{\BB_{a \By}\,\BG_{b \By}-\BG_{a \By}\,\BB_{b \By}}{\BG_{\By\By}}\,,\qquad 
 \AB_{a \Ay} = -\frac{\BG_{a \By}}{\BG_{\By\By}} \,, \qquad
 \Exp{2\Aphi}= \frac{\Exp{2\Bphi}}{\BG_{\By\By}} \,,
\\
 \AC_\Ay &= \BC_0 \,,\qquad 
 \AC_a = \BC_{a \By} - \BC_0\,\BB_{a \By}\,,\qquad 
 \AC_{ab \Ay} = \BC_{ab} - \frac{\BC_{a \By}\,\BG_{b \By}-\BG_{a \By}\,\BC_{b \By}}{\BG_{\By\By}} \,,
\\
 \AC_{abc} &=\BD_{abc \By} -\frac{3}{2}\,\epsilon_{\gamma\delta}\,\BB^\gamma_{[ab}\,\BB^\delta_{c]\By}
 - \frac{3\,\epsilon_{\gamma\delta}\,\BB^\gamma_{[a|\By|}\,\BB^\delta_{b|\By|}\,\BG_{c]\By}}{\BG_{\By\By}} 
\nn\\
 &=\BC_{abc \By} - 3 \, \BC_{[ab}\,\BB_{c]\By} - \frac{6\,\BC_{[a|\By|}\,\BB_{b|\By|}\,\BG_{c]\By}}{\BG_{\By\By}}\,,
\\
 \AC_{a_1\cdots a_4 \Ay} &=\BD_{a_1\cdots a_4} + 3 \BB_{[a_1a_2}\, \BC_{a_3a_4]}
\nn\\
 &\quad -\frac{4\,\BD_{[a_1a_2a_3|\By|}\,\BG_{a_4]\By}}{\BG_{\By\By}} - \frac{6 \BB_{[a_1|\By|}\,\BC_{a_2a_3}\,\BG_{a_4]\By}}{\BG_{\By\By}} -\frac{6 \BC_{[a_1|\By|} \BB_{a_2a_3}\,\BG_{a_4]\By}}{\BG_{\By\By}}
\nn\\
 &=\BC_{a_1\cdots a_4} -\frac{4\,\BC_{[a_1a_2a_3|\By|}\,\BG_{a_4]\By}}{\BG_{\By\By}} \,,
\\
 \AC_{a_1\cdots a_5} &= \BC_{a_1\cdots a_5} - 5\,\BD_{a_1\cdots a_4}\,\BB_{a_5\By} -15\,\BB_{[a_1a_2}\, \BC_{a_3a_4}\,\BB_{a_5]\By}
\nn\\
 &\quad -\frac{20\,\BD_{[a_1a_2a_3\By}\,\BB_{a_4|\By|}\,\BG_{a_5]\By}}{\BG_{\By\By}} + \frac{30\,\BB_{[a_1a_2}\,\BB_{a_3|\By|}\,\BC_{a_4|\By|}\,\BG_{a_5]\By}}{\BG_{\By\By}}
\nn\\
 &= \BC_{a_1\cdots a_5} -5\,\BC_{a_1\cdots a_4}\,\BB_{a_5\By} -\frac{20\,\BC_{[a_1a_2a_3\By}\,\BB_{a_4|\By|}\,\BG_{a_5]\By}}{\BG_{\By\By}} \,,
\\
 \AB_{a_1\cdots a_5 \Ay}&= \bbB_{a_1\cdots a_5 \By} +5\,\BD_{[a_1a_2a_3|\By|}\,\BC_{a_4a_5]} + \frac{5}{2}\, \epsilon_{\gamma\delta}\,\BC_{[a_1a_2}\,\BB^\gamma_{a_3a_4}\,\BB^\delta_{a_5]\By}
\nn\\
 &\quad - \frac{10\,\BD_{[a_1\cdots a_3|\By|}\,\BC_{a_4|\By|}\BG_{a_5]\By}}{\BG_{\By\By}} - \frac{15\, \BC_{[a_1a_2}\, \BB_{a_3|\By|}\,\BC_{a_4|\By|}\,\BG_{a_5]\By}}{\BG_{\By\By}} 
\nn\\
 &= \BB_{a_1\cdots a_5 \By} -5\,\BD_{[a_1\cdots a_4}\,\BC_{a_5]\By} -5\,\BD_{[a_1a_2a_3|\By|}\,\BC_{a_4a_5]} 
 - \frac{45}{2}\, \BC_{[a_1a_2}\,\BB_{a_3a_4}\,\BC_{a_5]\By} 
\nn\\
 &- \frac{15}{2}\, \BC_{[a_1a_2}\,\BC_{a_3a_4}\,\BB_{a_5]\By} - \frac{10\,\BD_{[a_1\cdots a_3|\By|}\,\BC_{a_4|\By|}\BG_{a_5]\By}}{\BG_{\By\By}} - \frac{15\, \BC_{[a_1a_2}\, \BB_{a_3|\By|}\,\BC_{a_4|\By|}\,\BG_{a_5]\By}}{\BG_{\By\By}}
\,. 
\end{align}

The above transformation rules completely match the transformation rules (under a $T$-duality along $x^\Ay$ direction) obtained in \cite{Eyras:1998hn} (see Appendix A therein) if we make the following identifications for the supergravity fields:
\begin{align}
 \begin{pmatrix} \mathsf{g}_{\mu\nu} \\ \cB \\ C^{(p+1)}\\ \tilde{\cB} \end{pmatrix}_{\!\!\!{\text{(IIA)}\atop\text{\cite{Eyras:1998hn}}}}
 =\begin{pmatrix}
 \AG_{pq}\\ \AB_2\\ \AC_{p+1}\\ -\AB_6 \end{pmatrix}_{\!\!\!{\text{(IIA)}\atop\text{here}}} ,\quad
 \begin{pmatrix}
 g_{\mu\nu}\\ B \\ C^{(2)}\\ C^{(4)}\\ C^{(6)}\\ \tilde{B}
\end{pmatrix}_{\!\!\!{\text{(IIB)}\atop\text{\cite{Eyras:1998hn}}}}
 = \begin{pmatrix}
 \BG_{pq}\\ \BB_2\\ -\BC_2\\ -\BD_4\\ -\bigl(\BC_6-\tfrac{1}{4}\,\BB_2\wedge \BB_2\wedge \BC_2\bigr) \\ -\bigl(\BB_6-\tfrac{1}{4}\,\BC_2\wedge \BC_2\wedge \BB_2\bigr) 
\end{pmatrix}_{{\!\!\!{\text{(IIB)}\atop\text{here}}}} .
\end{align}
As usual, the transformation rule for the Ramond-Ramond potentials can be summarized as
\begin{align}
\begin{split}
 \AC_{a_1\cdots a_{n-1}\Ay}&= \BC_{a_1\cdots a_{n-1}} - (n-1)\,\frac{\BC_{[a_1\cdots a_{n-2}|\By|}\,\BG_{a_{n-1}]\By}}{\BG_{\By\By}}\,,
\\
 \AC_{a_1\cdots a_n} &= \BC_{a_1\cdots a_n\By} - n\, \BC_{[a_1\cdots a_{n-1}}\, \BB_{a_n]\By} - n\,(n-1)\,\frac{\BC_{[a_1\cdots a_{n-2}|\By|}\, \BB_{a_{n-1}|\By|}\,\BG_{a_n]\By}}{\BG_{\By\By}}\,. 
\end{split}
\end{align}
If we define Ramond-Ramond poly-forms as
\begin{align}
 \AA \equiv \Exp{-\AB_2}\wedge\, \AC \,,\qquad 
 \BA \equiv \Exp{-\BB_2}\wedge\, \BC \qquad 
 \biggl(\AC \equiv \sum_{p\text{: odd}} \AC_p \,,\qquad 
 \BC \equiv \sum_{p\text{: even}} \BC_p\biggr) \,,
\end{align}
which appear in the D-brane worldvolume action, the $n$-form parts of $\AA$ and $\BA$, denoted by $\AA_n$ and $\BA_n$, satisfy a simple $T$-duality transformation rule \cite{Hassan:1999mm} (see Appendix \ref{app:conventions})
\begin{align}
 \AA_{a_1\cdots a_{n-1}\Ay} = \BA_{a_1\cdots a_{n-1}} \,,
\qquad
 \AA_{a_1\cdots a_n} = \BA_{a_1\cdots a_n\By} \,. 
\end{align}

\subsection{Buscher's rules for non-geometric potentials}

In the double field theory, instead of using the conventional supergravity fields $(G_{mn},\,B_{mn})$ we can also parameterize the generalized metric in terms of the dual fields $(\tilde{g}_{mn},\,\beta^{mn})$ \cite{Andriot:2011uh,Andriot:2012wx,Andriot:2012an,Blumenhagen:2012nk,Blumenhagen:2012nt,Blumenhagen:2013aia,Andriot:2013xca,Andriot:2014qla}. 
The familiar supergravity backgrounds can be described well by the former parameterization, but the latter non-geometric parameterization is necessary when we consider non-geometric backgrounds, such as $T$-folds or the non-Riemannian background studied in \cite{Lee:2013hma}. 
Therefore, both parameterizations are equally important. 
The non-geometric parameterization for the generalized metric in EFTs was considered in \cite{Malek:2012pw,Blair:2014zba} for the $\SL(5)$ EFT and in \cite{Lee:2016qwn} for $E_{d(d)}$ EFT ($d\leq 7$). 
Following the convention of \cite{Lee:2016qwn}, the non-geometric parameterization in the M-theory is given by
\begin{align}
 \cM_{IJ} &= (L_6^\rmT\,L_3^\rmT\,\hat{\cM}\,L_3\,L_6)_{IJ} \,,\quad L_3 =\Exp{\ell_3}\,,\quad L_6 = \Exp{\ell_6}\,,
\label{eq:non-geometric-M}
\\
 \hat{\cM} &\equiv \abs{\tilde{G}}^{\frac{1}{n-2}}\,
 \begin{pmatrix}
 \tilde{G}_{ij} & 0 & 0 & 0 \\
 0 & \tilde{G}^{i_1i_2,\,j_1j_2} & 0 & 0 \\
 0 & 0 & \tilde{G}^{i_1\cdots i_5,\,j_1\cdots j_5} & 0 \\
 0 & 0 & 0 & \tilde{G}^{i_1\cdots i_7,\,j_1\cdots j_7}\,\tilde{G}^{ij}
 \end{pmatrix}\,, \quad 
 \abs{\tilde{G}}\equiv \det(\tilde{G}_{ij})\,,
\\
 \ell_3 &\equiv {\footnotesize
 \begin{pmatrix}
 0 & -\frac{3!\,\delta^{i j_1j_2}_{l_1l_2l_3}}{\sqrt{2!}} & 0 & 0 \\
 0 & 0 & \frac{5!\,\delta_{i_1i_2 l_1l_2l_3}^{j_1\cdots j_5}}{\sqrt{2!\,5!}} & 0 \\
 0 & 0 & 0 & \frac{7!\,3\,\delta^{j_1\cdots j_7}_{i_1\cdots i_5 [l_1l_2}\,\delta^{j}_{l_3]}}{\sqrt{5!\,7!}} \\
 0 & 0 & 0 & 0
 \end{pmatrix}\,\frac{\Omega^{l_1l_2l_3}}{3!}}\,,
\\
 \ell_6 &\equiv {\footnotesize
 \begin{pmatrix}
 0 & 0 & -\frac{6!\,\delta^{i j_1\cdots j_5}_{l_1\cdots l_6}}{\sqrt{5!}} & 0 \\
 0 & 0 & 0 & \frac{7!\,6\,\delta^{j_1\cdots j_7}_{i_1i_2[l_1\cdots l_5}\, \delta^j_{l_6]}}{\sqrt{2!\,7!}} \\
 0 & 0 & 0 & 0 \\
 0 & 0 & 0 & 0
 \end{pmatrix}\,\frac{\Omega^{l_1\cdots l_6}}{6!}} \,.
\end{align}
On the other hand, the non-geometric parameterization in the type IIB theory is given by \cite{Lee:2016qwn}
\begin{align}
 &\sfM_{\sfM\sfN} = (\sfL_6^\rmT\,\sfL_4^\rmT\,\sfL_2^\rmT\, \hat{\sfM}\, \sfL_2\,\sfL_4\,\sfL_6)_{\sfM\sfN} \,,\quad \sfL_2\equiv \Exp{\ell_2}\,,\quad \sfL_4\equiv \Exp{\ell_4}\,,\quad \sfL_6\equiv \Exp{\ell_6}\,, 
\label{eq:non-geometric-IIB}
\\
 &\hat{\sfM} = \abs{\tilde{\BEG}}^{\frac{1}{n-2}}{\arraycolsep=0.5mm \left(\begin{array}{ccccc} 
 \tilde{\BEG}_{\sfm\sfn} & 0 & 0 & 0 & 0 \\
 0 & \tilde{m}_{\alpha\beta} \,\tilde{\BEG}^{\sfm\sfn} & 0 & 0 & 0 \\
 0 & 0 & \tilde{\BEG}^{\sfm_1\sfm_2\sfm_3,\,\sfn_1\sfn_2\sfn_3} & 0 & 0 \\
 0 & 0 & 0 & \tilde{m}_{\alpha\beta} \,\tilde{\BEG}^{\sfm_1\cdots \sfm_5,\,\sfn_1\cdots \sfn_5} & 0 \\
 0 & 0 & 0 & 0 & \tilde{\BEG}^{\sfm_1\cdots \sfm_6,\,\sfn_1\cdots \sfn_6}\,\tilde{\BEG}^{\sfm\sfn} \end{array}\right)} ,
\\
 &\bigl(\tilde{m}_{\alpha\beta}\bigr)= \Exp{\tilde{\Bphi}}\,\begin{pmatrix}
 \Exp{-2\tilde{\Bphi}} & \gamma^0 \\
 \gamma^0 & 1 + (\gamma^0)^2
 \end{pmatrix}\,, 
\\
 &\ell_2 = {\arraycolsep=0.5mm \left(\begin{array}{ccccc} 
 ~0~ & -2!\,\delta^\gamma_{\beta}\,\delta^{\sfm\sfn}_{\sfp_1\sfp_2} & 0 & 0 & 0 \\
 0 & 0 & \frac{3!\,\epsilon^{\alpha\gamma} \,\delta^{\sfn_1\sfn_2\sfn_3}_{\sfm\sfp_1\sfp_2}}{\sqrt{3!}} & 0 & 0 \\
 0 & 0 & 0 & \frac{5!\,\delta_\beta^\gamma\,\delta^{\sfn_1\cdots\sfn_5}_{\sfm_1\sfm_2\sfm_3\sfp_1\sfp_2}\,}{\sqrt{3!\,5!}} & 0 \\
 0 & 0 & 0 & 0 & -\frac{6!\,2\,\epsilon^{\alpha\gamma}\,\delta_{\sfm_1\cdots\sfm_5[\sfp_1}^{\sfn_1\cdots\sfn_6}\,\delta_{\sfp_2]}^\sfn}{\sqrt{5!\,6!}} \\
 0 & 0 & 0 & 0 & 0 
 \end{array}\right)\,\frac{\beta_\gamma^{\sfp_1\sfp_2}}{2!}}\,,
\\
 &\ell_4 = {\arraycolsep=0.5mm \left(\begin{array}{ccccc} 
 ~0~ & ~0~ & -\frac{4!\,\delta^{\sfm \sfn_1\sfn_2\sfn_3}_{\sfp_1\cdots\sfp_4}}{\sqrt{3!}} & 0 & 0 \\
 0 & 0 & 0 & -\frac{5!\,\delta^\alpha_\beta \,\delta^{\sfn_1\cdots\sfn_5}_{\sfm\sfp_1\cdots\sfp_4}}{\sqrt{5!}} & 0 \\
 0 & 0 & 0 & 0 & -\frac{6!\,4\,\delta_{\sfm_1\sfm_2\sfm_3[\sfp_1\sfp_2\sfp_3}^{\sfn_1\cdots \sfn_6}\,\delta_{\sfp_4]}^\sfn}{\sqrt{3!\,6!}} \\
 0 & 0 & 0 & 0 & 0 \\
 0 & 0 & 0 & 0 & 0 
 \end{array}\right)\frac{\eta^{\sfp_1\cdots\sfp_4}}{4!}}\,, 
\\
 &\ell_6= {\arraycolsep=0.5mm \left(\begin{array}{ccccc} 
 ~0~ & ~0~ & ~0~ & -\frac{6!\,\delta^\gamma_\beta\,\delta^{\sfm\sfn_1\cdots \sfm_5}_{\sfp_1\cdots\sfp_6}}{\sqrt{5!}} & 0 \\
 0 & 0 & 0 & 0 & \frac{6!\,6\,\epsilon^{\alpha\gamma}\,\delta_{\sfm [\sfp_1\cdots \sfp_5}^{\sfn_1\cdots \sfn_6}\,\delta^{\sfn}_{\sfp_6]}}{\sqrt{6!}} 
\\
 0 & 0 & 0 & 0 & 0 \\
 0 & 0 & 0 & 0 & 0 \\
 0 & 0 & 0 & 0 & 0
 \end{array}\right)\,\frac{\eta^{\sfp_1\cdots\sfp_6}_\gamma}{6!} } \,.
\end{align}

Note that we can easily obtain the identifications of non-geometric potentials between the M-theory and the type IIB theory without repeating a similar calculation as that performed above. 
In fact, the parameterization \eqref{eq:non-geometric-M} for $\cM_{IJ}$ can be obtained by calculating the inverse, $\cM^{IJ}\equiv (\cM^{-1})^{IJ}$, of \eqref{eq:M-theory-conventional} and making the replacements,
\begin{align}
 G^{ij}\to \tilde{G}_{ij}\,, \quad A_{i_1i_2i_3}\to -\Omega^{i_1i_2i_3}\,,\quad A_{i_1\cdots i_6}\to -\Omega^{i_1\cdots i_6} \,,\quad 
 \cM^{IJ}\to \cM_{IJ} \,. 
\label{eq:replacements-M}
\end{align}
Similarly, the parameterization \eqref{eq:non-geometric-IIB} for $\sfM_{\sfM\sfN}$ is also obtained by calculating the inverse, $\sfM^{\sfM\sfN}\equiv (\sfM^{-1})^{\sfM\sfN}$ of \eqref{eq:IIB-conventional} and making the replacements,
\begin{align}
\begin{split}
 &\BEG^{\sfm\sfn}\to \tilde{\BEG}_{\sfm\sfn}\,, \quad \Bphi\to \tilde{\Bphi}\,, \quad \BC_0\to -\gamma^0\,, \quad \BB^\alpha_{\sfm\sfn}\to -\beta^{\sfm\sfn}_\alpha\,,
\\
 &\BD_{\sfm_1\cdots \sfm_4}\to -\eta^{\sfm_1\cdots \sfm_4}\,,\quad \BD^\alpha_{\sfm_1\cdots \sfm_6}\to -\eta^{\sfm_1\cdots \sfm_6}_\alpha \,,\quad 
 \sfM^{\sfM\sfN} \to \sfM_{\sfM\sfN}\,. 
\end{split}
\label{eq:replacements-IIB}
\end{align}
Then, using
\begin{align}
 \bigl(S^{-1} \cM^{-1}\,S^{-\rmT}\bigr)^{\sfM\sfN}
 = (S^{-1})^\sfM{}_I\,(S^{-\rmT})^J{}_\sfN\, \cM^{IJ}= \sfM^{\sfM\sfN} \,,
\end{align}
which follows from \eqref{eq:linear-map}, and making the replacements \eqref{eq:replacements-M} and \eqref{eq:replacements-IIB}, we obtain
\begin{align}
 (\tilde{S}^{-1})^\sfM{}_I\,(\tilde{S}^{-\rmT})^J{}_\sfN\, \tilde{\cM}_{IJ}= \tilde{\sfM}_{\sfM\sfN} \,,\qquad 
 (\tilde{S}^{-\rmT})^I{}_\sfM \equiv (S^{-\rmT})_I{}^\sfM \,,
\end{align}
where $\tilde{\cM}_{IJ}$ and $\tilde{\sfM}_{\sfM\sfN}$ are the same as \eqref{eq:non-geometric-M} and \eqref{eq:non-geometric-IIB}, respectively, and the tilde is added just in order to clarify that it is parameterized with the non-geometric potentials. 
We can easily show that $(\tilde{S}^{-\rmT})^I{}_\sfM =S^I{}_\sfM$ and the matrix $S^I{}_\sfM$ connects $\tilde{\cM}_{IJ}$ and $\tilde{\sfM}_{\sfM\sfN}$. 

From the identifications \eqref{eq:replacements-M} and \eqref{eq:replacements-IIB}, if we define
\begin{align}
\begin{split}
 \tilde{\BG}_{\sfm\sfn} &\equiv \Exp{\frac{1}{2}\,\tilde{\Bphi}}\tilde{\BEG}_{\sfm\sfn} \,,\qquad
 \gamma^{\sfm_1\cdots \sfm_4}\equiv \eta^{\sfm_1\cdots \sfm_4} - 3\,\beta^{[\sfm_1\sfm_2}\, \gamma^{\sfm_3\sfm_4]} \,,
\\
 \gamma^{\sfm_1\cdots \sfm_6}
 &\equiv \eta_1^{\sfm_1\cdots \sfm_6}-15\,\eta^{[\sfm_1\cdots \sfm_4}\,\beta^{\sfm_5\sfm_6]} - 15\, \beta^{[\sfm_1\sfm_2}\,\beta^{\sfm_3\sfm_4}\, \gamma^{\sfm_5\sfm_6]}\,,
\\
 \beta^{\sfm_1\cdots \sfm_6}
 &\equiv \eta_2^{\sfm_1\cdots \sfm_6} - 15\,\eta^{[\sfm_1\cdots \sfm_4}\,\gamma^{\sfm_5\sfm_6]} +30\,\gamma^{[\sfm_1\sfm_2}\,\gamma^{\sfm_3\sfm_4}\,\beta^{\sfm_5\sfm_6]}\,,
\end{split}
\end{align}
and decompose the eleven-dimensional fields into the fields in the type IIA theory,
\begin{align}
\begin{split}
 \tilde{G}^{rs} &= \Exp{-\frac{2}{3}\,\tilde{\Aphi}}\,\tilde{\AG}^{rs}+\Exp{\frac{4}{3}\,\Aphi}\,\Agamma^r\, \Agamma^s \,,\quad 
 \tilde{G}^{r \Az} = -\Exp{\frac{4}{3}\,\tilde{\Aphi}}\, \Agamma^r \,,
\\
 \tilde{G}^{\Az\Az} &= \Exp{\frac{4}{3}\,\tilde{\Aphi}}\,,\quad
 \Omega^{pq \Az} = \Abeta^{pq} \,,\quad 
 \Omega^{pqr} = \Agamma^{pqr} \,,
\\
 \Omega^{p_1\cdots p_5\Az} &= \Agamma^{p_1\cdots p_5} - 5\,\Agamma^{[p_1p_2p_3}\,\Abeta^{p_4p_5]} \,,\quad 
 \Omega^{p_1\cdots p_6} = \Abeta^{p_1\cdots p_6} \,,
\end{split}
\end{align}
we obtain the following Buscher's rules for non-geometric potentials:
\begin{align}
\begin{split}
 \tilde{\AG}^{ab} &= \tilde{\BG}^{ab} - \frac{\tilde{\BG}^{a \By}\,\tilde{\BG}^{b \By}-\beta^{a \By}\,\beta^{b \By}}{\tilde{\BG}^{\By\By}}\,,\quad 
 \tilde{\AG}^{a \Ay}= \frac{\beta^{a \By}}{\tilde{\BG}^{\By\By}}\,,\quad 
 \tilde{\AG}^{\Ay\Ay}=\frac{1}{\tilde{\BG}^{\By\By}}\,,
\\
 \Abeta^{ab} &= \beta^{ab} + \frac{\beta^{a \By}\,\tilde{\BG}^{b \By}-\tilde{\BG}^{a \By}\,\beta^{b \By}}{\tilde{\BG}^{\By\By}}\,,\quad 
 \Abeta^{a \Ay} = \frac{\tilde{\BG}^{a \By}}{\tilde{\BG}^{\By\By}} \,, \quad
 \Exp{-2\Aphi}= \frac{\Exp{-2\Bphi}}{\tilde{\BG}^{\By\By}} \,,
\\
 \Agamma^{a_1\cdots a_{n-1}\Ay}&= \gamma^{a_1\cdots a_{n-1}} - (n-1)\,\frac{\gamma^{[a_1\cdots a_{n-2}|\By|}\,\tilde{\BG}^{a_{n-1}]\By}}{\tilde{\BG}^{\By\By}}\,,
\\
 \Agamma^{a_1\cdots a_n} &= \gamma^{a_1\cdots a_n \By} + n\, \gamma^{[a_1\cdots a_{n-1}}\, \beta^{a_n]\By} + n\,(n-1)\,\frac{\gamma^{[a_1\cdots a_{n-2}|\By|}\, \beta^{a_{n-1}|\By|}\,\tilde{\BG}^{a_n] \By}}{\tilde{\BG}^{\By\By}}\,,
\\
 \Abeta^{a_1\cdots a_5 \Ay}&= \beta^{a_1\cdots a_5 \By} + 5\,\eta^{[a_1\cdots a_4}\,\gamma^{a_5]\By} + 5\,\eta^{[a_1a_2a_3|\By|}\,\gamma^{a_4a_5]} 
 - \frac{45}{2}\, \gamma^{[a_1a_2}\,\beta^{a_3a_4}\,\gamma^{a_5]\By} 
\\
 &- \frac{15}{2}\, \gamma^{[a_1a_2}\,\gamma^{a_3a_4}\,\beta^{a_5]\By} + \frac{10\,\eta^{[a_1\cdots a_3|\By|}\,\gamma^{a_4|\By|}\tilde{\BG}^{a_5]\By}}{\tilde{\BG}^{\By\By}} - \frac{15\, \gamma^{[a_1a_2}\, \beta^{a_3|\By|}\,\gamma^{a_4|\By|}\,\tilde{\BG}^{a_5]\By}}{\tilde{\BG}^{\By\By}}
\,,
\end{split}
\end{align}
where the inverse of the string-frame metric is denoted by $\tilde{\BG}^{\sfm\sfn}\equiv (\tilde{\BG}^{-1})^{\sfm\sfn}$\,. 
If we define the wedge product for $p$-vectors, $v=(1/p!)\,v^{i_1\cdots i_p}\,\partial_{i_1}\vee \cdots \vee \partial_{i_p}$, in the same manner as $p$-forms, and define poly-vectors,
\begin{align}
\begin{split}
 \Aalpha &\equiv \Exp{\Abeta^2}\vee\, \Agamma \quad
 \biggl(\Agamma \equiv \!{\textstyle\sum\limits_{p\text{: odd}}}\! \Agamma^p \,,\quad 
 \Agamma^p \equiv \tfrac{1}{p!}\,\Agamma^{\sfm_1\cdots\sfm_p}\,\partial_{\sfm_1}\vee \partial_{\sfm_p}\,,\quad 
 \Abeta^2 \equiv \tfrac{1}{2!}\,\beta^{\sfm\sfn}\,\partial_\sfm\vee \partial_\sfn\biggr) \,,
\\
 \alpha &\equiv \Exp{\beta^2}\vee\, \gamma \quad
 \biggl(\gamma \equiv \!{\textstyle\sum\limits_{p\text{: even}}}\! \gamma^p \,,\quad 
 \gamma^p \equiv \tfrac{1}{p!}\,\gamma^{\sfm_1\cdots\sfm_p}\,\partial_{\sfm_1}\vee \partial_{\sfm_p}\,,\quad 
 \beta^2 \equiv \tfrac{1}{2!}\,\beta^{\sfm\sfn}\,\partial_\sfm\vee \partial_\sfn\biggr) \,,
\end{split}
\end{align}
their $n$-vector parts, $\Aalpha^n$ and $\alpha^n$, obey a simple $T$-duality transformation rule,
\begin{align}
 \Aalpha^{a_1\cdots a_n} = \alpha^{a_1\cdots a_n\By}\,,\qquad 
 \Aalpha^{a_1\cdots a_{n-1}\Ay} = \alpha^{a_1\cdots a_{n-1}} \,,
\label{eq:alpha-rule}
\end{align}
similar to the modified Ramond-Ramond forms $\AA_n$ and $\BA_n$\,. 
Note that the non-geometric $P$-fluxes defined in \cite{Lee:2016qwn} can be expressed as
\begin{align}
\begin{split}
 \mathcal{P} &\equiv \rmd \Agamma + \rmd\Abeta\vee \Agamma 
 = \Exp{-\Abeta}\vee\,\rmd \bigl(\Exp{\Abeta}\vee\,\Agamma\bigr) 
 = \Exp{-\Abeta}\vee\,\rmd \Aalpha\,,
\\
 \mathsf{P} &\equiv \rmd \gamma + \rmd\beta\vee \gamma 
 = \Exp{-\beta}\vee\,\rmd \bigl(\Exp{\beta}\vee\,\gamma\bigr) 
 = \Exp{-\beta}\vee\,\rmd \alpha\,,
\end{split}
\end{align}
where $\vee$ is defined to commute with the exterior derivative (see section 4.2 of \cite{Andriot:2014qla} where the above expression has been conjectured). 

A family of co-dimension 2 branes, called an exotic $p^{7-p}_3$-brane ($0\leq p\leq 7$) (see \cite{Obers:1998fb} for the notation), is known to be a magnetic source of the $P$-flux \cite{Bergshoeff:2010xc,Bergshoeff:2011zk,Bergshoeff:2011se,Andriot:2013xca,Chatzistavrakidis:2013jqa,Kimura:2014upa,Andriot:2014qla,Sakatani:2014hba}. 
In particular, the $7_3$-brane, which is also known as the NS7-brane, is the $S$-dual of the D7-brane and it magnetically couples to the $\gamma^0$ potential in the same way as the D7-brane magnetically couples to the $C_0$ potential. 
On the other hand, an object which electrically couples to the $C_0$ potential is known as the D-instanton or the D$(-1)$-brane \cite{Gibbons:1995vg}, and similarly an object which electrically couples to the $\gamma^0$ potential will also be an instanton, to be called the $\overline{\text{D}(-1)}$-brane, whose mass is proportional to the string coupling constant $g_s$ (bound states of these instantons are known as Q-instantons \cite{Bergshoeff:2008qq}). 
Under multiple $T$-duality transformations, the $\overline{\text{D}(-1)}$-brane is mapped to another object. 
From the transformation rule \eqref{eq:alpha-rule}, the potential $\gamma^0 = \alpha^0$ is also mapped to an $n$-vector $\Aalpha^{a_1\cdots a_n}$ (or $\alpha^{a_1\cdots a_n}$), and the object can be identified as the electric source of $\Aalpha^{a_1\cdots a_n}$. 
Since $\Aalpha^{a_1\cdots a_n}$ is a 0-form, such an object will also be an instanton, as discussed in \cite{Bergshoeff:2011se,Sakatani:2014hba}. 
A similar object which couples to the $\beta$-field has been studied recently in \cite{Sakatani:2014hba,Blair:2015eba,Arvanitakis:2016zes,Blair:2016xnn}, and it will be interesting to study these instantons in more detail. 

\section{Summary and Outlook}
\label{sec:summary}

In this note, we made a connection between the M-theory parameterization and the type IIB parameterization for $E_{d(d)}$ EFT with $d\leq 7$, and obtained the duality transformation rules for non-geometric potentials in EFT. 
Our linear map will be useful when we concretely apply various results in EFT to the M-theory or the type IIB theory. 

We can generalize the same calculation to the $E_{8(8)}$ EFT. 
In this case, the decomposition of the $248$ generalized coordinates (suitable for the M-theory parameterization) is given by \cite{West:2004kb,Godazgar:2013rja,Strickland-Constable:2013xta,Hohm:2014fxa}
\begin{align*}
 (x^I) = (\underbrace{x^i_{\vphantom{o}}}_{\mathrm{P}\ [8]},\,\underbrace{y_{i_1i_2}}_{\mathrm{M}2\ [28]},\,\underbrace{y_{i_1\cdots i_5}}_{\mathrm{M}5\ [56]},\,\underbrace{y_{i_1\cdots i_7,\,j}}_{\mathrm{KKM}/8\ [63]},\,\underbrace{y_{i_1\cdots i_8}}_{8\ [1]},\, \underbrace{y_{i_1\cdots i_8,\,j_1j_2j_3}}_{5^3\ [56]},\,\underbrace{y_{i_1\cdots i_8,\,j_1\cdots j_6}}_{2^6\ [28]},\,\underbrace{y_{i_1\cdots i_8,\,j_1\cdots j_8,\,k}}_{0^{(1,7)}\ [8]}) \,,
\end{align*}
and each coordinate is associated with a brane charge as specified under the brace. 
The numbers in square brackets represent the number of independent winding charges, which sum to 248. 
Here, note that $y_{i_1\cdots i_7,\,j}$ satisfies $y_{[i_1\cdots i_7,\,j]}=0$. 
The branes that correspond to 56 coordinates, $y_{i_1\cdots i_7,\,j}$ with $j\in \{i_1,\dotsc,i_7\}$, are the Kaluza-Klein monopoles (KKM). 
On the other hand, the remaining eight branes, corresponding to $y_{i_1\cdots i_7,\,j}$ with $j\not\in \{i_1,\dotsc ,i_7\}$ and $y_{i_1\cdots i_8}$, may be the unfamiliar 8-branes discussed in \cite{Elitzur:1997zn,Hull:1997kb,Obers:1998fb}, whose mass is given by
\begin{align}
 M= \frac{R_3\cdots R_{10}}{\ell_{11}^9} \,,
\end{align}
where $R_i$ is the radius of the $x^i$ coordinate on the 8-torus and $\ell_{11}$ is the eleven-dimensional Planck length. 

In terms of the type IIB theory, the decomposition of the $248$ generalized coordinates is given by \cite{Riccioni:2007ni,Strickland-Constable:2013xta,Hohm:2014fxa}
\begin{align*}
 (x^\sfM) = &(\underbrace{x^\sfm_{\vphantom{o}}}_{\mathrm{P}\ [7]},\,\underbrace{\sfy^\alpha_\sfm}_{\mathrm{F}1/\mathrm{D}1\ [7\times 2]},\,\underbrace{\sfy_{\sfm_1\sfm_2\sfm_3}}_{\mathrm{D}3\ [35]},\,\underbrace{\sfy^\alpha_{\sfm_1\cdots \sfm_5}}_{\mathrm{NS}5/\mathrm{D}5\ [21\times 2]},\,\underbrace{\sfy_{\sfm_1\cdots \sfm_6,\,\sfn}}_{\mathrm{KKM}/7_2\ [48]},\,\underbrace{\sfy_{\sfm_1\cdots \sfm_7}}_{7_2\ [1]},\, \underbrace{\sfy^{\alpha\beta}_{\sfm_1\cdots \sfm_7}}_{\mathrm{Q}7\ [3]}, 
\nn\\
 &\, \underbrace{\sfy^\alpha_{\sfm_1\cdots \sfm_7,\,\sfn_1\sfn_2}}_{5^2_2/5^2_3\ [21\times 2]},\, \underbrace{\sfy_{\sfm_1\cdots \sfm_7,\,\sfn_1\cdots \sfn_4}}_{3^4_3\ [35]},\, \underbrace{\sfy^\alpha_{\sfm_1\cdots \sfm_7,\,\sfn_1\cdots \sfn_6}}_{1^6_4/1^6_3\ [7\times 2]},\, \underbrace{\sfy_{\sfm_1\cdots \sfm_7,\,\sfn_1\cdots \sfn_7,\,\mathsf{p}}}_{0_4^{(1,6)}\ [7]}) \,,
\end{align*}
where $\sfy_{[\sfm_1\cdots \sfm_6,\,\sfn]}=0$ and $\sfy^{[\alpha\beta]}_{\sfm_1\cdots \sfm_7}=0$\,. 
Again, branes which correspond to 42 coordinates, $\sfy_{\sfm_1\cdots \sfm_6,\,\sfn}$ with $\sfn\in \{\sfm_1,\dotsc ,\sfm_6\}$, are the Kaluza-Klein monopoles (KKM), while seven branes, corresponding to $\sfy_{\sfm_1\cdots \sfm_6,\,\sfn}$ with $\sfn\not\in \{\sfm_1,\dotsc ,\sfm_6\}$ and $\sfy_{\sfm_1\cdots \sfm_7}$, are $7_2$-branes \cite{Obers:1998fb} with mass
\begin{align}
 M= \frac{R_3\cdots R_9}{g_s^2\,l_s^8} \,,
\label{eq:7_2-mass}
\end{align}
where $l_s$ is the string length. 
As there are eight branes with the same mass in the M-theory, there will be one more $7_2$-brane, namely a brane with the mass \eqref{eq:7_2-mass}, in the type IIB theory. 
A natural expectation (see also \cite{West:2004st}) is that the $7_2$-brane together with the D7-brane and $7_3$-brane (or the NS7-brane) constitute a triplet of 7-branes which corresponds to $\sfy^{\alpha\beta}_{\sfm_1\cdots \sfm_7}$, known as the Q7-brane \cite{Meessen:1998qm,Bergshoeff:2002mb,Bergshoeff:2007aa} (see also \cite{Meessen:1998qm,DallAgata:1998ahf,Bergshoeff:2005ac,Bergshoeff:2007aa} for the 8-form potentials $A^{\alpha\beta}_{\sfm_1\cdots \sfm_8}$ which couple to the Q7-brane). 

In the $E_{8(8)}$ EFT, from the tensor structure, we can consider the identifications between the generalized coordinates given in Table \ref{eq:identify-E8}. 
There, $c_1,\dotsc,c_6$ are constants to be determined by requiring that the $(d-2)$-dimensional tensors are reorganized into $(d-1)$-dimensional tensors in the type IIB theory. 
A small calculation suggests $c_1=c_3=c_4=c_5=c_6=1$, and a complete analysis will be performed in future work. 
\begin{table}[t]
\begin{align*}
\footnotesize
 \begin{pmatrix}
 \sfx^a \\[-2mm] \sfx^{\By} \\ \hline 
 \sfy^\alpha_a \\[-2mm] \sfy^\alpha_{\By} \\ \hline
 \sfy_{a_1a_2a_3} \\[-2mm] \sfy_{a_1a_2 \By} \\ \hline \sfy^\alpha_{a_1\cdots a_5} \\[-2mm] \sfy^\alpha_{a_1\cdots a_4\By} \\ \hline 
 \tilde{\sfy}_{a_1\cdots a_5\By,b} \\[-2mm] 
 \sfy_{a_1\cdots a_5\By,\By} \\[-2mm] 
 \sfy_{a_1\cdots a_6,b} \\[-2mm] 
 \sfy_{a_1\cdots a_6,\By} \\ \hline
 \sfy_{a_1\cdots a_6\By}\\ \hline 
 \sfy^{\alpha\beta}_{a_1\cdots a_6\By}\\ \hline 
 \sfy^\alpha_{a_1\cdots a_6\By,b_1b_2}\\[-2mm]
 \sfy^\alpha_{a_1\cdots a_6\By,b \By}\\ \hline
 \sfy_{a_1\cdots a_6\By,b_1\cdots b_4}\\[-2mm]
 \sfy_{a_1\cdots a_6\By,b_1b_2b_3\By}\\ \hline
 \sfy^\alpha_{a_1\cdots a_6\By,b_1\cdots b_6}\\[-2mm]
 \sfy^\alpha_{a_1\cdots a_6\By,b_1\cdots b_5\By}\\ \hline
 \sfy_{a_1\cdots a_6\By,b_1\cdots b_6\By,c}\\[-2mm]
 \sfy_{a_1\cdots a_6\By,b_1\cdots b_6\By,\By}
 \end{pmatrix} 
 = \begin{pmatrix}
 x^a \\[-2mm]
 y_{\Ay\Az} \\ \hline
 \epsilon^{\alpha\beta}\,y_{a\beta} \\[-2mm]
 x^\alpha \\ \hline
 y_{a_1a_2a_3\Ay\Az} \\[-2mm] 
 y_{a_1a_2} \\ \hline
 \epsilon^{\alpha\beta}\,y_{a_1\cdots a_5 \Ay\Az,\beta} \\[-2mm] 
 \epsilon^{\alpha\beta}\,y_{a_1\cdots a_4\beta} \\ \hline
 \tilde{y}_{a_1\cdots a_5 \Ay\Az,b} \\[-2mm]
 y_{a_1\cdots a_5} \\[-2mm]
 c_1\,y_{a_1\cdots a_6\Ay\Az,b\Ay\Az} \\[-2mm]
 c_2\,y_{a_1\cdots a_6[\Ay,\Az]} \\ \hline
 y_{a_1\cdots a_6\Ay\Az}\\ \hline 
 \epsilon^{\alpha\gamma}\epsilon^{\beta\delta}\,y_{a_1\cdots a_6(\gamma,\delta)}\\ \hline 
 \epsilon^{\alpha\beta}\,y_{a_1\cdots a_6\Ay\Az,b_1b_2\beta} \\[-2mm]
 c_3\, \epsilon^{\alpha\beta}\,y_{a_1\cdots a_6\beta,b}\\ \hline
 y_{a_1\cdots a_6\Ay\Az,b_1\cdots b_4\Ay\Az}\\[-2mm]
 c_4\, y_{a_1\cdots a_6\Ay\Az,b_1b_2b_3} \\ \hline
 \epsilon^{\alpha\beta}\,y_{a_1\cdots a_6\Ay\Az,b_1\cdots b_6\Ay\Az,\beta} \\[-2mm]
 c_5\,\epsilon^{\alpha\beta}\,y_{a_1\cdots a_6\Ay\Az,b_1\cdots b_5\beta}\\ \hline
 y_{a_1\cdots a_6\Ay\Az,b_1\cdots b_6\Ay\Az,c} \\[-2mm]
 c_6\,y_{a_1\cdots a_6\Ay\Az,b_1\cdots b_6}
 \end{pmatrix}
 \quad \text{or}\quad 
 \begin{pmatrix}
 x^a \\[-2mm] x^\alpha \\[0.9mm] \hline 
 y_{a_1a_2} \\[-2mm] y_{a \alpha} \\[-2mm] 
 y_{\Ay\Az} \\[0.9mm] \hline y_{a_1\cdots a_5} \\[-2mm] 
 y_{a_1\cdots a_4\beta} \\[-2mm] 
 y_{a_1a_2a_3 \Ay\Az} \\[0.9mm] \hline 
 \tilde{y}_{a_1\cdots a_5\Ay\Az,b} \\[-2mm]
 y_{a_1\cdots a_5 \Ay\Az,\alpha}\\[-2mm]
 y_{a_1\cdots a_6\alpha,b} \\[-2mm]
 y_{a_1\cdots a_6(\alpha,\beta)} \\[-2mm]
 y_{a_1\cdots a_6[\Ay,\Az]} \\[0.9mm] \hline
 y_{a_1\cdots a_6\Ay\Az} \\[0.9mm] \hline
 y_{a_1\cdots a_6\Ay\Az,b_1b_2b_3}\\[-2mm]
 y_{a_1\cdots a_6\Ay\Az,b_1b_2\alpha}\\[-2mm]
 y_{a_1\cdots a_6\Ay\Az,b\Ay\Az}\\[0.9mm] \hline
 y_{a_1\cdots a_6\Ay\Az,b_1\cdots b_6}\\[-2mm]
 y_{a_1\cdots a_6\Ay\Az,b_1\cdots b_5\alpha}\\[-2mm]
 y_{a_1\cdots a_6\Ay\Az,b_1\cdots b_4\Ay\Az}\\[0.9mm] \hline
 y_{a_1\cdots a_6\Ay\Az,b_1\cdots b_6\Ay\Az,c}\\[-2mm]
 y_{a_1\cdots a_6\Ay\Az,b_1\cdots b_6\Ay\Az,\alpha} 
 \end{pmatrix}
 =
\begin{pmatrix}
 \sfx^a \\[-2mm]
 \sfy_\By^\alpha \\[0.9mm] \hline
 \sfy_{a_1a_2\By} \\[-2mm]
 \epsilon^\rmT_{\alpha\beta}\,\sfy_a^\beta \\[-2mm]
 \sfx^\By \\[0.9mm] \hline
 \sfy_{a_1\cdots a_5\By,\By} \\[-2mm]
 \epsilon^\rmT_{\alpha\beta}\,\sfy^\beta_{a_1\cdots a_4} \\[-2mm]
 \sfy_{a_1a_2a_3} \\[0.9mm] \hline
 \tilde{\sfy}_{a_1\cdots a_5\By,b} \\[-2mm]
 \epsilon^\rmT_{\alpha\beta}\,\sfy_{a_1\cdots a_5}^\beta \\[-2mm]
 c_3^{-1}\,\epsilon^\rmT_{\alpha\beta}\,\sfy^\beta_{a_1\cdots a_6\By,b\By}\\[-2mm]
 \epsilon_{\alpha\gamma}\,\epsilon_{\beta\delta}\,\sfy^{\gamma\delta}_{a_1\cdots a_6\By} \\[-2mm]
 c_2^{-1}\,\sfy_{a_1\cdots a_6,\By}\\[0.9mm] \hline
 \sfy_{a_1\cdots a_6\By} \\[0.9mm] \hline
 c_4^{-1}\,\sfy_{a_1\cdots a_6\By,b_1b_2b_3\By}\\[-2mm]
 \epsilon^\rmT_{\alpha\gamma}\,\sfy^\gamma_{a_1\cdots a_6\By,b_1b_2}\\[-2mm]
 c_1^{-1}\,\sfy_{a_1\cdots a_6,b}\\[0.9mm] \hline
 c_6^{-1}\,\sfy_{a_1\cdots a_6\By,b_1\cdots b_6\By,\By} \\[-2mm]
 c_5^{-1}\,\epsilon^\rmT_{\alpha\gamma}\,\sfy^\gamma_{a_1\cdots a_6\By,b_1\cdots b_5\By} \\[-2mm]
 \sfy_{a_1\cdots a_6\By,b_1\cdots b_4}\\[0.9mm] \hline
 \sfy_{a_1\cdots a_6\By,b_1\cdots b_6\By,c}\\[-2mm]
 \epsilon^\rmT_{\beta\gamma}\,\sfy^\gamma_{a_1\cdots a_6\By,b_1\cdots b_6} 
\end{pmatrix} \,.
\end{align*}
\caption{A conjectural map for the $E_{8(8)}$ EFT. We have defined $\tilde{\sfy}_{a_1\cdots a_5\By,b}\equiv \sfy_{a_1\cdots a_5\By,b}-\sfy_{[a_1\cdots a_5|\By,|b]}$ and $\tilde{y}_{a_1\cdots a_5 \Ay\Az,b}\equiv y_{a_1\cdots a_5 \Ay\Az,b} -y_{[a_1\cdots a_5| \Ay\Az,|b]}$.}
\label{eq:identify-E8}
\end{table}

The following are some applications of our results:
\begin{itemize}
\item If we have a solution of the eleven-dimensional supergravity that depends on eleven coordinates $x^i$, by substituting the background fields into the generalized metric $\cM_{IJ}(x^i)$, $\cM_{IJ}(x^i)$ satisfies the equations of motion of EFT. 
In principle, we can read off the type IIB fields from the same generalized metric $\cM_{IJ}(x^i)$, and at the same time, we can also rename the coordinates $x^i$ as $(x^a,\,\sfy_{\By}^\alpha)$ using the linear map \eqref{eq:map-coordinates}. 
This is still a solution of EFT, but it has dual-coordinate dependence, much like the localized monopole solution discussed in \cite{Jensen:2011jna,Berman:2014jsa}. 
Thus, it is not a solution of the conventional type IIB supergravity. 
Further studies on such non-geometric solutions in EFT will be interesting. 

\item According to the idea proposed in \cite{Hull:2004in,Asakawa:2012px}, it may be possible to interpret a D-brane as a $(1+9)$-dimensional surface (called the Dirac manifold) embedded into a doubled spacetime with coordinates $(x^m,\,\tilde{x}_m)$ ($m=0,\dotsc,9$)\,. 
If the $(1+9)$-dimensional surface is extending in the $(x^a,\,\tilde{x}_i)$ ($a=0,\dotsc, p$, $i=p+1,\dotsc,9$) directions and localized in the $(x^i,\,\tilde{x}_a)$ directions, it is called a D$p$-brane, since it has a $(p+1)$-dimensional worldvolume in the physical spacetime. 
We can generalize this idea to the exceptional spacetime by using our linear map. 
If we consider a 9-brane extending in the $(\sfx^0,\dotsc,\sfx^9)$ directions, after a $T$-duality along the $\sfx^9$ direction (i.e.~our linear map, $\sfx^9\to y_{9\Az}$, etc.), we obtain a brane extending in the $(x^0,\dotsc,x^8,y_{9\Az})$ directions. 
Further taking a $T$-duality along the $x^8$ direction, we obtain a brane extending along the $(\sfx^0,\dotsc,\sfx^7,\sfy^1_8,\sfy^1_9)$ directions. 
Repeating the $T$-duality transformations, we obtain various branes extending in the following directions:
\begin{align}
\begin{cases}
 (\sfx^0,\,\dotsc,\,\sfx^{p},\,\sfy^1_{p+1},\dotsc,\,\sfy^1_9) & (p:\text{odd})
\\
 (x^0,\,\dotsc,\,x^{p},\,y_{(p+1)\Az},\dotsc,\,y_{9\Az}) & (p:\text{even})
\end{cases}\,. 
\end{align}
As we can see from the parameterization of the generalized metric, $y_{i\Az}$ in the type IIA theory or $\sfy^1_\sfm$ in the type IIB theory can be identified with the dual coordinates $\tilde{x}_m$ in the double field theory. 
Therefore, following the idea of \cite{Hull:2004in,Asakawa:2012px}, the above brane may be interpreted as a D$p$-brane. 
In EFT, we can proceed further since we also have the $S$-duality transformation. 
In the following, we consider various rotations of the same 9-brane in the exceptional spacetime; let us call the brane a $(\cdots)$-brane if the brane is extending in the $(\cdots)$ directions. 
Performing the $S$-duality transformation, a $(\sfx^0,\,\dotsc,\,\sfx^5,\,\sfy^1_6,\dotsc,\,\sfy^1_9)$-brane (D5-brane) is mapped to a $(\sfx^0,\,\dotsc,\,\sfx^5,\,\sfy^2_6,\dotsc,\,\sfy^2_9)$-brane, which will be interpreted as the type IIB NS5-brane. 
Further performing a $T$-duality along the $\sfx^6$ direction, we obtain a $(x^0,\,\dotsc,\,x^5,\,x^\Az,\,y_{67},y_{68},\,y_{69})$-brane, which may be the Kaluza-Klein monopole in the M-theory, where $x^6$ will be the Taub-NUT direction. 
If we swap the M-theory circle $x^\Az$ and the Taub-NUT direction, we obtain a D6-brane, $(x^0,\,\dotsc,\,x^6,\,y_{7\Az},y_{8\Az},\,y_{9\Az})$. 
A $(\sfx^0,\,\dotsc,\,\sfx^5,\,\sfy^1_6,\sfy^1_7,\,\sfy_{678},\,\sfy_{679})$-brane may be interpreted as the exotic $5^2_2$-brane. 
Exchanging $x^2$ and $x^\Az$ in the D2-brane, $(x^0,\,x^1,\,x^2,\,y_{3\Az},\dotsc,\,y_{9\Az})$, we obtain a $(x^0,\,x^1,\,x^\Az,\,y_{23},\dotsc,\,y_{29})$-brane, which may be the type IIA fundamental string. 
If this viewpoint works well, various known branes may be interpreted as a single object that is wrapping on various dual directions. 
This viewpoint is consistent with the proposal that \textit{strings and branes are waves} propagating in various dual directions \cite{Berkeley:2014nza,Berman:2014jsa,Berman:2014hna}. 
It will be interesting to develop this viewpoint further by finding a $U$-duality invariant action for the brane in the exceptional spacetime, and also by showing how various brane tensions can be reproduced. 
\end{itemize}

\section*{Acknowledgment}
Y.S.~would like to thank Soo-Jong Rey and Kanghoon Lee for useful discussions and collaborations on related topics. 

\appendix

\section{Conventions}
\label{app:conventions}

In this appendix, we summarize our conventions.

\subsection{Conventions for supergravity fields}

We clarify our conventions for the supergravity fields by displaying various expressions. 

In the eleven-dimensional supergravity, the gauge transformations of the gauge potentials can be read off from the parameterization \eqref{eq:M-theory-conventional}:
\begin{align}
 \delta A_3 = \rmd \lambda_2\,,\quad \delta A_6 = \rmd \lambda_5 + \frac{1}{2}\, \rmd \lambda_2\wedge A_3\,. 
\end{align}
We define the invariant field strengths,
\begin{align}
 F_4 = \rmd A_3\,,\quad F_7 = \rmd A_6 + \frac{1}{2}\,A_3\wedge F_4 \,,
\end{align}
which satisfy the Bianchi identity,
\begin{align}
 \rmd F_4 =0 \,,\quad \rmd F_7 - \frac{1}{2}\,F_4 \wedge F_4 =0 \,. 
\end{align}
If we also consider diffeomorphisms, the gauge symmetries become
\begin{align}
 \delta A_3 = \Lie_v A_3 + \rmd \lambda_2\,,\quad 
 \delta A_6 = \Lie_v A_6 + \rmd \lambda_5 + \frac{1}{2}\, \rmd \lambda_2\wedge A_3 \,. 
\end{align}

Considering the dimensional reduction to the type IIA theory, the gauge potentials and the field strengths are decomposed as
\begin{align}
\begin{split}
 A_3 &= \AC_3 + \AB_2\wedge \rmd x^\Az \,,\quad F_4 = G_4 + H_3 \wedge (\rmd x^\Az + \AC_1) \,,
\\
 A_6 &= \AB_6 + \bigl(\AC_5 - \tfrac{1}{2}\, \AC_3\wedge \AB_2\bigr)\wedge \rmd x^\Az \,, \quad
 F_7 = H_7 + G_6 \wedge (\rmd x^\Az + \AC_1) \,,
\end{split}
\end{align}
where we defined the field strengths,
\begin{align}
\begin{split}
 G_6 &\equiv \rmd \AC_5 - H_3 \wedge \AC_3\,,\quad 
 G_4 \equiv \rmd \AC_3 - H_3 \wedge \AC_1\,,\quad 
 G_2 \equiv \rmd \AC_1 \,,
\\
 H_7 &\equiv \rmd \AB_6 - G_6 \wedge \AC_1 + \frac{1}{2}\,G_4\wedge \AC_3 - \frac{1}{2}\,H_3 \wedge \AC_3 \wedge \AC_1 \,,\quad 
 H_3 \equiv \rmd \AB_2 \,,
\end{split}
\end{align}
which satisfy the Bianchi identities
\begin{align}
\begin{split}
 &\rmd G_6 - H_3\wedge G_4 = 0\,,\quad 
 \rmd G_4 - H_3\wedge G_2 = 0\,,\quad 
 \rmd G_2 = 0\,,
\\
 &\rmd H_7 + G_6\wedge G_2 - \frac{1}{2}\,G_4 \wedge G_4 = 0 \,,\quad 
 \rmd H_3=0 \,. 
\end{split}
\end{align}
From the gauge symmetries in the eleven-dimensional supergravity, we obtain the gauge symmetries in the type IIA supergravity,
\begin{align}
\begin{split}
 \delta \AC_1 &= \Lie_v \AC_1 + \rmd \lambda_0 \,,
\\
 \delta \AB_2 &= \Lie_v \AB_2 + \rmd \omega_1 \,,
\\
 \delta \AC_3 &= \Lie_v \AC_3 + \rmd \lambda_2 + \rmd \lambda_0\wedge \AB_2 \,,
\\
 \delta \AC_5 &= \Lie_v \AC_5 + \rmd \lambda_4 + \rmd \lambda_2\wedge \AB_2 + \frac{1}{2}\,\rmd \lambda_0\wedge \AB_2\wedge \AB_2 \,,
\\
 \delta \AB_6 &= \Lie_v \AB_6 + \rmd \omega_5 + \frac{1}{2}\,\rmd \lambda_2\wedge\AC_3 - \rmd \lambda_0\wedge \bigl(\AC_5-\tfrac{1}{2}\,\AC_3\wedge \AB_2\bigr) \,.
\end{split}
\end{align}

The gauge symmetries in the type IIB supergravity, with our convention \eqref{eq:IIB-conventional}, become
\begin{align}
\begin{split}
 \delta \BB^\alpha_2 &= \Lie_v \BB^\alpha_2 + \rmd \lambda^\alpha_1 \,,
\\
 \delta \BD_4 &= \Lie_v \BD_4 + \rmd \lambda_3 + \frac{1}{2}\,\epsilon_{\gamma\delta}\,\rmd\lambda_1^\gamma\wedge \BB^\delta_2 \,,
\\
 \delta \BD^\alpha_6 &= \Lie_v \BD^\alpha_6 + \rmd \lambda^\alpha_5 - \rmd \lambda^\alpha_1\wedge \BD_4 - \frac{1}{6}\, \epsilon_{\gamma\delta}\,\rmd\lambda_1^\gamma\wedge \BB^\delta_2\wedge \BB^\alpha_2 \,. 
\end{split}
\end{align}
These are equivalent to
\begin{align}
\begin{split}
 \delta \BB_2 &= \Lie_v\BB_2+\rmd \omega_1\quad (\omega_1\equiv\lambda^1_1) \,,
\\
 \delta \BC_2 &= \Lie_v\BC_2+\rmd \lambda_1\quad (\lambda_1\equiv -\lambda^2_1) \,,
\\
 \delta \BC_4 &= \Lie_v\BC_4+\rmd \lambda_3 + \rmd\lambda_1\wedge \BB_2 \,,
\\
 \delta \BC_6 &= \Lie_v\BC_6+\rmd \lambda_5 +\rmd \lambda_3\wedge \BB_2 + \frac{1}{2}\, \rmd \lambda_1 \wedge \BB_2\wedge\BB_2\quad (\lambda_5\equiv\lambda^1_5)\,,
\\
 \delta \BB_6 &= \Lie_v\BB_6+\rmd \omega_5 +\rmd \lambda_3\wedge \BC_2 + \rmd \lambda_1 \wedge \BB_2\wedge\BC_2\quad (\omega_5\equiv -\lambda^2_5)\,.
\end{split}
\end{align}
The field strengths in the type IIB supergravity are defined by
\begin{align}
\begin{split}
 G_7&\equiv \rmd \BC_6 - H_3 \wedge \BC_4\,,\quad 
 G_5 \equiv \rmd \BC_4 - H_3 \wedge \BC_2\,,\quad
 G_3 \equiv \rmd \BC_2 - H_3 \wedge \BC_0\,,\quad
 G_1 \equiv \rmd \BC_0 \,, 
\\
 H_7 &\equiv \rmd \BB_6 - \BC_4\wedge\rmd \BC_2 -\frac{1}{2}\,\BC_2\wedge\BC_2\wedge H_3 - \BC_0 \,G_7 \,,\quad H_3 \equiv \rmd \BB_2 \,,
\end{split}
\end{align}
and these satisfy the Bianchi identities,
\begin{align}
\begin{split}
 &\rmd G_7 - H_3 \wedge G_5=0\,,\quad 
 \rmd G_5 - H_3 \wedge G_3=0\,,\quad 
 \rmd G_3 - H_3 \wedge G_1=0\,,\quad 
 \rmd G_1 =0\,, 
\\
 &\rmd H_7 + G_1\wedge G_7 + G_5 \wedge G_3 = 0 \,,\quad \rmd H_3 =0 \,.
\end{split}
\end{align}
If we define poly-forms,
\begin{align}
 G_{\text{IIA/IIB}}\equiv \sum_{p\text{: even/odd}} G_p\,,\qquad \lambda_{\text{IIA/IIB}} \equiv \sum_{p\text{: even/odd}} \lambda_p \,,
\end{align}
we obtain
\begin{align}
\begin{split}
 \delta \AC = \Lie_v \AC + \rmd \lambda_{\text{IIA}} \wedge \Exp{\AB_2}\,,\quad G_{\text{IIA}}=\rmd \AC - H_3\wedge \AC\,,\quad \rmd G_{\text{IIA}} - H_3\wedge G_{\text{IIA}} =0 \,, 
\\
 \delta \BC = \Lie_v \BC + \rmd \lambda_{\text{IIB}} \wedge \Exp{\BB_2}\,,\quad G_{\text{IIB}}=\rmd \BC - H_3\wedge \BC\,,\quad \rmd G_{\text{IIB}} - H_3\wedge G_{\text{IIB}} =0 \,. 
\end{split}
\end{align}
A convenient expression for the $T$-duality rule is as follows. 
If we consider a $T$-duality along the $\sfx^\By$ direction in a type IIB background,
\begin{align}
\begin{split}
 &\rmd \mathsf{s}^2 = \widetilde{\BG}_{ab}\,\rmd \sfx^a\,\rmd \sfx^b +\BG_{\By\By}\,\bigl(\rmd \sfx^\By+\mathsf{a}_1\bigr)^2\,,\quad \Exp{2\Bphi}\,,
\\
 &\BB_2 = \widetilde{\BB}_2 + \mathsf{b}_1\wedge \bigl(\rmd \sfx^\By+\mathsf{a}_1\bigr)\,, \quad
 \BC = \widetilde{\BC} + \Bc\wedge \bigl(\rmd \sfx^\By+\mathsf{a}_1\bigr) 
\\
 &\bigl(\mathsf{a}_1,\, \mathsf{b}_1:\text{1-forms}\,,\quad \widetilde{\BB}_2:\text{2-form}\,,\quad \widetilde{\BC}:\text{even poly-form}\,,\quad \Bc :\text{odd poly-form}\bigr)\,,
\end{split}
\end{align}
we obtain the following dual background on the type IIA side,
\begin{align}
\begin{split}
 \rmd \mathfrak{s}^2&= \widetilde{\BG}_{ab}\,\rmd x^a\,\rmd x^b + \BG^{-1}_{\By\By} \,\bigl(\rmd x^\Ay-\mathsf{b}_1\bigr)^2\,,\quad 
 \Exp{2\Aphi} = \frac{\Exp{2\Bphi}}{\BG_{\By\By}} \,,
\\
 \AB_2 &= \widetilde{\BB}_2 - \mathsf{a}_1\wedge \rmd x^\Ay \,, \quad
 \AC = \Bc + \widetilde{\BC} \wedge \bigl(\rmd x^\Ay - \mathsf{b}_1\bigr) \,.
\end{split}
\end{align}
Using this rule, we can show that modified Ramond-Ramond potentials $\BA\equiv \Exp{-\BB_2}\BC$ and $\AA\equiv \Exp{-\AB_2}\AC$ follow a simple transformation rule. 
Indeed, in the original type IIB background, we have
\begin{align}
 \BA &= \Exp{-\BB_2}\BC = \Exp{-\widetilde{\BB}_2} \bigl(\widetilde{\BC} +\Bc\wedge \mathsf{a}_1- \widetilde{\BC} \wedge \mathsf{b}_1\wedge \mathsf{a}_1\bigr) + \Exp{-\widetilde{\BB}_2}\bigl(\Bc - \widetilde{\BC} \wedge \mathsf{b}_1\bigr)\wedge \rmd \sfx^\By \,,
\end{align}
while $\AA$ in the transformed background becomes
\begin{align}
 \AA =\Exp{-\AB_2}\AC = \Exp{-\widetilde{\BB}_2} \bigl(\Bc -\widetilde{\BC}\wedge \mathsf{b}_1\bigr) + \Exp{-\widetilde{\BB}_2} \bigl(\widetilde{\BC} + \Bc \wedge \mathsf{a}_1 -\widetilde{\BC}\wedge \mathsf{b}_1\wedge \mathsf{a}_1 \bigr)\wedge \rmd \sfx^\Ay \,. 
\end{align}
Namely, the transformation rule is simply given by
\begin{align}
 \AA_{a_1\cdots a_p} = \BA_{a_1\cdots a_p\By}\,,\qquad 
 \AA_{a_1\cdots a_p\Ay} = \BA_{a_1\cdots a_p}\,. 
\end{align}

\subsection{Period of generalized coordinates}

We choose the periods of the physical coordinates in the M-theory and the type IIB theory as
\begin{align}
 x^i\sim x^i+2\pi R^i\,,\qquad 
 \sfx^\sfm\sim \sfx^\sfm+2\pi \BR^\sfm\,,
\end{align}
where the radius of the M-theory circle is identified as $R^\Az = g_s\,l_s$\,. 
The Planck length in eleven dimensions is given by $\ell_{11}=g_s^{1/3}\,l_s$\,. 
From our linear map and the transformation rules,
\begin{align}
\begin{split}
 &\text{$T$-duality}:\quad R \to l_s^2/R \,, \quad 
 g_s \to g_s\, l_s/R \,,\quad
 l_s \to l_s \,,
\\
 &\text{$S$-duality}:\quad g_s \to 1/g_s \,, \quad 
 l_s = g_s^{1/2}\, l_s \,,
\end{split}
\end{align}
we can postulate the following radius in the dual directions:
\begin{align}
\begin{split}
 &\underline{\text{M-theory:}}
\\
 &\quad R^{y_{i_1i_2}} = \frac{\ell_{11}^3}{R^{i_1} R^{i_2}}\,,\quad 
 R^{y_{i_1\cdots i_5}} = \frac{\ell_{11}^6}{R^{i_1}\cdots R^{i_5}}\,,\quad 
 R^{y_{i_1\cdots i_7,\,j}} = \frac{\ell_{11}^9}{R^{i_1}\cdots R^{i_7}R^j}\,,\\
 &\underline{\text{IIB-theory:}}
\\
 &\quad \BR^{\sfy_\sfm^\alpha} = 
\begin{pmatrix}
 \frac{l_s^2}{\BR^{\sfm}} \\ \frac{g_s\,l_s^2}{\BR^{\sfm}}
\end{pmatrix}
\,,\quad 
 \BR^{\sfy_{\sfm_1\sfm_2\sfm_3}} = \frac{g_s\,l_s^4}{\BR^{\sfm_1} \BR^{\sfm_2} \BR^{\sfm_3}}\,,\quad 
 \BR^{\sfy_{\sfm_1\cdots\sfm_5}^\alpha} = \begin{pmatrix}
 \frac{g_s\,l_s^6}{\BR^{\sfm_1} \cdots \BR^{\sfm_5}} \\ \frac{g_s^2\,l_s^6}{\BR^{\sfm_1} \cdots \BR^{\sfm_5}}
\end{pmatrix}\,,
\\
 &\quad \BR^{\sfy_{\sfm_1\cdots \sfm_6,\,\sfm}} = \frac{g_s^2\,l_s^8}{\BR^{\sfm_1}\cdots \BR^{\sfm_6}\BR^{\sfm}}\,. 
\end{split}
\end{align}

\section{Outline of the computation}
\label{app:outline}

Here, we explain a way to check the linear map \eqref{eq:linear-map} and \eqref{eq:M-IIB-map}. 

\subsection{M-theory: decomposition into $(d-2)+2$ dimensions}

If we consider, for example, the combination,
\begin{align}
 \frac{v_{i_1\cdots i_p}}{\sqrt{p!}}\,A^{i_1\cdots i_p,\,j_1\cdots j_p}\,\frac{w_{j_1\cdots j_p}}{\sqrt{p!}} \qquad
 \bigl(A^{i_1\cdots i_p,\,j_1\cdots j_p} 
 = A^{[i_1\cdots i_p],\,j_1\cdots j_p} = A^{i_1\cdots i_p,\,[j_1\cdots j_p]}\bigr)\,,
\end{align}
under the $(d-2)+2$ dimensions, it can be expanded as
\begin{align}
 &\begin{pmatrix}
 \frac{v_{a_1\cdots a_p}}{\sqrt{p!}}
 & \frac{v_{a_1\cdots a_{p-1}\alpha}}{\sqrt{(p-1)!}}
 & \frac{v_{a_1\cdots a_{p-2}\Ay\Az}}{\sqrt{(p-2)!}}
 \end{pmatrix}
\nn\\
 &{\footnotesize\times {\arraycolsep=0.5mm \left(\begin{array}{ccc}
 A^{a_1\cdots a_p,\,b_1\cdots b_p}
 & \sqrt{p}\,A^{a_1\cdots a_p,\,b_1\cdots b_{p-1}\beta}
 & \sqrt{p(p-1)}\,A^{a_1\cdots a_p,\,b_1\cdots b_{p-2}\Ay\Az} 
\\
 \sqrt{p}\,A^{a_1\cdots a_{p-1}\alpha,\,b_1\cdots b_p}
 & p\,A^{a_1\cdots a_{p-1}\alpha,\,b_1\cdots b_{p-1}\beta}
 & p\sqrt{p-1}\,A^{a_1\cdots a_{p-1}\alpha,\,b_1\cdots b_{p-2}\Ay\Az}
\\
 \sqrt{p(p-1)}\,A^{a_1\cdots a_{p-2}\Ay\Az,\,b_1\cdots b_p}
 & p\sqrt{p-1}\,A^{a_1\cdots a_{p-2}\Ay\Az,\,b_1\cdots b_{p-1}\beta}
 & p(p-1)\,A^{a_1\cdots a_{p-2}\Ay\Az,\,b_1\cdots b_{p-2}\Ay\Az}
 \end{array}\right)} \begin{pmatrix}
 \frac{w_{b_1\cdots b_p}}{\sqrt{p!}} \\
 \frac{w_{b_1\cdots b_{p-1}\beta}}{\sqrt{(p-1)!}} \\
 \frac{w_{b_1\cdots b_{p-2}\Ay\Az}}{\sqrt{(p-2)!}} 
 \end{pmatrix}.}
\end{align}
Following the rule, we decompose a generalized vector as
\begin{align}
 (V^I) = \begin{pmatrix}
 v^i \\ \frac{v_{i_1i_2}}{\sqrt{2!}}\\
 \frac{v_{i_1\cdots i_5}}{\sqrt{5!}}\\
 \frac{v_{i_1\cdots i_7,\,i}}{\sqrt{7!}}
\end{pmatrix}
 = {\footnotesize\begin{pmatrix}
 v^a \\[-2mm] v^\alpha \\ \hline \frac{v_{a_1a_2}}{\sqrt{2!}} \\[-2mm]
 v_{a \alpha} \\[-2mm] v_{\Ay\Az} \\ \hline
 \frac{v_{a_1\cdots a_5}}{\sqrt{5!}}\\[-2mm]
 \frac{v_{a_1\cdots a_4\alpha}}{\sqrt{4!}}\\[-2mm]
 \frac{v_{a_1a_2a_3,\,\Ay\Az}}{\sqrt{3!}}\\ \hline
 \frac{v_{a_1\cdots a_5\Ay\Az,\,a}}{\sqrt{5!}} \\[-2mm]
 \frac{v_{a_1\cdots a_5\Ay\Az,\,\alpha}}{\sqrt{5!}}
 \end{pmatrix}} \,,
\end{align}
and correspondingly decompose the generalized metric as follows:
\begin{align}
\hat{\cM} \equiv &\abs{G}^{\frac{1}{n-2}}
 {\footnotesize
 {\arraycolsep0.5mm \left(\begin{array}{cc|ccc|ccc|cc}
 G_{ab} & G_{a\beta} & 0 & 0 & 0 & 0 & 0 & 0 & 0 & 0 \\
 G_{\alpha b} & G_{\alpha\beta} & 0 & 0 & 0 & 0 & 0 & 0 & 0 & 0 \\ 
\hline
 0 & 0 & G^{a_1a_2,b_1b_2} & \sqrt{2}\,G^{a_1a_2,b\beta} & 
\sqrt{2}\,G^{a_1a_2,\Ay\Az} & 0 & 0 & 0 & 0 & 0 \\
 0 & 0 & \sqrt{2}\,G^{a \alpha,b_1b_2} & 2\,G^{a \alpha,b \beta} & 
2\,G^{a \alpha,\Ay\Az} & 0 & 0 & 0 & 0 & 0 \\
 0 & 0 & \sqrt{2}\,G^{\Ay\Az,b_1b_2} & 2\,G^{\Ay\Az,b \beta} & 
2\,G^{\Ay\Az,\Ay\Az} & 0 & 0 & 0 & 0 & 0 \\ \hline
 0 & 0 & 0 & 0 & 0 & G^{(5)}_1 &G^{(5)}_2 &G^{(5)}_3 & 0 & 0 \\
 0 & 0 & 0 & 0 & 0 & G^{(5)}_4 &G^{(5)}_5 &G^{(5)}_6 & 0 & 0 \\
 0 & 0 & 0 & 0 & 0 & G^{(5)}_7 &G^{(5)}_8 &G^{(5)}_9 & 0 & 0 
\\\hline
 0 & 0 & 0 & 0 & 0 & 0 & 0 & 0 & G^{(7,1)}_1 & G^{(7,1)}_2 \\
 0 & 0 & 0 & 0 & 0 & 0 & 0 & 0 & G^{(7,1)}_3 & G^{(7,1)}_4
 \end{array}\right) }\,, }
\\[5pt]
&{\footnotesize
 \arraycolsep=0.5mm \left(\begin{array}{ccc}
 G^{(5)}_1 &G^{(5)}_2 &G^{(5)}_3\\
 G^{(5)}_4 &G^{(5)}_5 &G^{(5)}_6\\
 G^{(5)}_7 &G^{(5)}_8 &G^{(5)}_9\\
\end{array}\right) \equiv\left(\begin{array}{ccc}
 G^{a_1\cdots a_5,b_1\cdots b_5}
 &\sqrt{5}\,G^{a_1\cdots a_5,b_1\cdots b_4\beta}
 &2\sqrt{5}\,G^{a_1\cdots a_5,b_1b_2b_3\Ay\Az}\\
 \sqrt{5}\,G^{a_1\cdots a_4\alpha,b_1\cdots b_5}
 &5\,G^{a_1\cdots a_4\alpha,b_1\cdots b_4\beta}
 &10\,G^{a_1\cdots a_4\alpha,b_1b_2b_3\Ay\Az}\\
 2\sqrt{5}\,G^{a_1a_2a_3\Ay\Az,b_1\cdots b_4\beta}
 &10\,G^{a_1a_2a_3\Ay\Az,b_1\cdots b_4\beta}
 &20\,G^{a_1a_2a_3\Ay\Az,b_1b_2b_3\Ay\Az} 
 \end{array}\right)} \,,
\\
&{\footnotesize
 {\arraycolsep=0.5mm \left(\begin{array}{cc}
 G^{(7,1)}_1 & G^{(7,1)}_2 \\ G^{(7,1)}_3 & G^{(7,1)}_4 \end{array}\right)
 \equiv 42\,G^{a_1\cdots a_5\Ay\Az,b_1\cdots b_5\Ay\Az}\,
 \left(\begin{array}{cc}
 G^{ab} & G^{a\beta} \\
 G^{\alpha b} & G^{\alpha\beta}
 \end{array}\right) }}\,,
\\
\ell_3 \equiv &{\footnotesize
 {\arraycolsep=0.5mm \left(\begin{array}{cc|ccc|ccc|cc}
 0 & 0 & 0 & 0 & 0 & 0 & 0 & 0 & 0 & 0 \\
 0 & 0 & 0 & 0 & 0 & 0 & 0 & 0 & 0 & 0 \\ \hline
 -\frac{A_{a_1a_2 b}}{\sqrt{2!}} & -\frac{A_{a_1a_2 \beta}}{\sqrt{2!}}
 & 0 & 0 & 0 & 0 & 0 & 0 & 0 & 0 \\
 -A_{a\alpha b} & - \epsilon_{\alpha\beta} A_{a \Ay\Az} & 0 & 0 & 0 
& 0 & 0 & 0 & 0 & 0 \\
 -A_{\Ay\Az b} & 0 & 0 & 0 & 0 & 0 & 0 & 0 & 0 & 0 \\ \hline
 0 & 0 & \frac{5!\delta^{b_1b_2c_1c_2c_3}_{a_1\cdots a_5}
 A_{c_1c_2c_3}}{3!\sqrt{2!\,5!}} & 0 & 0 & 0 & 0 & 0 & 0 & 0 
\\
 0 & 0 & \frac{4!\delta^{b_1b_2c_1c_2}_{a_1\cdots a_4}
 A_{c_1c_2\alpha}}{2!\sqrt{2!\,4!}}
 & \frac{4!\delta_\alpha^\beta\delta^{c_1c_2c_3 b}_{a_1\cdots 
a_4}
 A_{c_1c_2c_3}}{3!\sqrt{4!}} & 0 & 0 & 0 & 0 & 0 & 0 \\
 0 & 0 & \frac{3!\delta^{b_1b_2c}_{a_1a_2a_3}
 A_{c\Ay\Az}}{\sqrt{2!\,3!}}
 & \frac{3!\epsilon^{\beta\gamma} \delta^{bc_1c_2}_{a_1a_2a_3}
 A_{c_1c_2\gamma}}{2!\sqrt{3!}}
 & \frac{A_{a_1a_2a_3}}{\sqrt{3!}} & 0 & 0 & 0 & 0 & 0 \\ \hline
 0 & 0 & 0 & 0 & 0 &l^{(3)}_1 & l^{(3)}_2& l^{(3)}_3 & 0 & 0 \\
 0 & 0 & 0 & 0 & 0 & l^{(3)}_4 & l^{(3)}_5& l^{(3)}_6 & \,0\, & 0
 \end{array}\right) }}\,,
\\
 &{\footnotesize\arraycolsep=0.5mm \left(\begin{array}{ccc}
 l^{(3)}_1 & l^{(3)}_2& l^{(3)}_3 \\
 l^{(3)}_4 & l^{(3)}_5& l^{(3)}_6 \\
 \end{array}\right)
 \equiv
 \left(\begin{array}{ccc}
 -\delta^{b_1\cdots b_5}_{a_1\cdots a_5}A_{a \Ay\Az}
 &-\frac{5! \epsilon^{\beta\gamma}
 \delta^{b_1\cdots b_4c}_{a_1\cdots 
a_5}A_{ca\gamma}}{\sqrt{4!\,5!}}
 & -\frac{5! \delta^{b_1b_2b_3c_1c_2}_{a_1\cdots a_5} A_{a 
c_1c_2}}{2!\sqrt{3!\,5!}}\\
 0& -\frac{5!\delta_\alpha^\beta
 \delta^{b_1\cdots 
b_4c}_{a_1\cdots{}a_5}A_{c\Ay\Az}}{\sqrt{4!\,5!}}
 & -\frac{5! \delta^{b_1b_2b_3c_1c_2}_{a_1\cdots a_5} A_{c_1c_2 
\alpha}}{2!\sqrt{3!\,5!}}
\end{array}\right)}\,,
\\
 \ell_6 \equiv &{\footnotesize
 {\arraycolsep=0.5mm \left(\begin{array}{cc|ccc|ccc|cc}
 0 & 0 & 0 & 0 & 0 & 0 & 0 & 0 & 0 & 0 \\
 0 & 0 & 0 & 0 & 0 & 0 & 0 & 0 & 0 & 0 \\ \hline
 0 & 0 & 0 & 0 & 0 & 0 & 0 & 0 & 0 & 0 \\
 0 & 0 & 0 & 0 & 0 & 0 & 0 & 0 & 0 & 0 \\
 0 & 0 & 0 & 0 & 0 & 0 & 0 & 0 & 0 & 0 \\ \hline
 \frac{A_{a_1\cdots a_5 b}}{\sqrt{5!}} & \frac{A_{a_1\cdots a_5 \beta}}{\sqrt{5!}}
 & 0 & 0 & 0 & 0 & 0 & 0 & 0 & 0 \\
 \frac{A_{a_1\cdots a_4 \alpha b}}{\sqrt{4!}} & 
 \frac{\epsilon_{\alpha\beta} A_{a_1\cdots a_4 \Ay\Az}}{\sqrt{4!}}
 & 0 & 0 & 0 & 0 & 0 & 0 & 0 & 0 \\
 \frac{A_{a_1a_2a_3 \Ay\Az b}}{\sqrt{3!}} & 0
 & 0 & 0 & 0 & 0 & 0 & 0 & 0 & 0 \\ \hline
 0 & 0 & l^{(6)}_1 & l^{(6)}_2& l^{(6)}_3 & 0 & 0 & 0 & 0 & 0 \\
 0 & 0 & l^{(6)}_4 & l^{(6)}_5& l^{(6)}_6 & \,0\, & 0 & \,0\, & 
\,0\, & 0
 \end{array}\right) }}\,,\\
&{\footnotesize\arraycolsep=0.5mm \left(\begin{array}{ccc}
 l^{(6)}_1 & l^{(6)}_2& l^{(6)}_3 \\
 l^{(6)}_4 & l^{(6)}_5& l^{(6)}_6 \\
 \end{array}\right)
 \equiv
 \left(\begin{array}{ccc}
 -\frac{5!\delta^{b_1b_2c_1c_2c_3}_{a_1\cdots a_5} A_{a c_1c_2c_3\Ay\Az}}{3!\sqrt{2!\,5!}}
 & -\frac{5!\epsilon^{\beta\gamma} \delta^{bc_1\cdots c_4}_{a_1\cdots a_5} A_{a c_1\cdots c_4 \gamma}}{4!\sqrt{5!}} & 0\\
 0 & -\frac{5!\delta_\alpha^\beta \delta^{bc_1\cdots c_4}_{a_1\cdots a_5} A_{c_1\cdots c_4\Ay\Az}}{4!\sqrt{5!}}
 & -\frac{A_{\alpha a_1\cdots a_5}}{\sqrt{5!}}
\end{array}\right)}\,. 
\end{align}

\subsection{Type IIB theory: decomposition into $(d-2)+1$ dimensions}

As in the case of the M-theory, we decompose a generalized vector in the type IIB section as
\begin{align}
 (V^\sfM)
 = \begin{pmatrix}
 v^\sfm \\
 v_\sfm^\alpha \\
 \frac{v_{\sfm_1\sfm_2\sfm_3}}{\sqrt{3!}}\\
 \frac{v_{\sfm_1\cdots \sfm_5}^\alpha}{\sqrt{5!}}\\
 \frac{v_{\sfm_1\cdots \sfm_7,\,\sfm}}{\sqrt{7!}}
\end{pmatrix}
 = \begin{pmatrix}
 v^a \\[-2mm]
 v^{\By} \\ \hline
 v^\alpha_a \\[-2mm]
 v^\alpha_{\By} \\ \hline
 \frac{v_{a_1a_2a_3} }{\sqrt{3!}}\\[-2mm]
 \frac{v_{a_1a_2 \By} }{\sqrt{2!}}\\ \hline
 \frac{v^\alpha_{a_1\cdots a_5} }{\sqrt{5!}}\\[-2mm]
 \frac{v^\alpha_{a_1\cdots a_4\By}}{\sqrt{4!}}\\ \hline
 \frac{v_{a_1\cdots a_6\By,\,a}}{\sqrt{6!}}\\[-2mm]
 \frac{v_{a_1\cdots a_6\By,\,\By}}{\sqrt{6!}}
 \end{pmatrix} \,,
\end{align}
and also the generalized metric as
\begin{align}
 \sfM_{\sfM\sfN}
 = (\Exp{\ell_3^\rmT}\Exp{\ell_2^\rmT}\Exp{\ell_1^\rmT}\,
 \hat{\sfM}\, \Exp{\ell_1} \Exp{\ell_2}\Exp{\ell_3} 
)_{\sfM\sfN} \,,
\label{eq:IIB-decomposed}
\end{align}
where
\begin{align}
 \hat{\sfM} ={} &{\footnotesize \Exp{-\frac{4}{n-2}\,\Bphi} 
\abs{\BG}^{\frac{1}{n-2}}
 {\arraycolsep=1.0mm\left(\begin{array}{cc|cc|cc|cc|cc}
 \BG_{ab} & \BG_{a\By} & 0 & 0 & 0 & 0 & 0 & 0 & 0 & 0 \\
 \BG_{\By b} & \BG_{\By\By} & 0 & 0 & 0 & 0 & 0 & 0 & 0 & 0 \\ \hline
 0 & 0 & \Exp{\Bphi}m_{\alpha\beta}\,\BG^{ab}
 &\Exp{\Bphi}m_{\alpha\beta}\,\BG^{a\By} & 0 & 0 & 0 & 0 & 0 & 0 
\\
 0 & 0 & \Exp{\Bphi}m_{\alpha\beta}\,\BG^{\By b}
 & \Exp{\Bphi}m_{\alpha\beta}\,\BG^{\By\By} & 0 & 0 & 0 & 0 & 0 & 
0 \\ \hline
 0 & 0 & 0 & 0 & \BG^{(3)}_1 & \BG^{(3)}_2 & 0 & 0 & 0 & 0 \\
 0 & 0 & 0 & 0 & \BG^{(3)}_3 & \BG^{(3)}_4 & 0 & 0 & 0 & 0 \\ \hline
 0 & 0 & 0 & 0 & 0 & 0 & \BG^{(5)}_1 & \BG^{(5)}_2 & 0 & 0 \\
 0 & 0 & 0 & 0 & 0 & 0 & \BG^{(5)}_3 & \BG^{(5)}_4 & 0 & 0 \\ \hline
 0 & 0 & 0 & 0& 0 & 0 & 0 & 0 & \BG^{(6,1)}_1 & \BG^{(6,1)}_2 \\
 0 & 0 & 0 & 0& 0 & 0 & 0 & 0 & \BG^{(6,1)}_3 & \BG^{(6,1)}_4 
 \end{array}\right)}} ,
\\
 &{\footnotesize {\arraycolsep=0.5mm
 \left(\begin{array}{cc} \BG^{(3)}_1 & \BG^{(3)}_2\\ \BG^{(3)}_3 & \BG^{(3)}_4 \end{array}\right)
 \equiv \left(\begin{array}{cc}
 \Exp{2\Bphi}\BG^{a_1a_2a_3,\,b_1b_2b_3}
 & \sqrt{3}\Exp{2\Bphi}\BG^{a_1a_2a_3,\,b_1b_2\By}\\
 \sqrt{3}\Exp{2\Bphi}\BG^{a_1a_2\By,\,b_1b_2b_3}
 & 3\Exp{2\Bphi}\BG^{a_1a_2\By,\,b_1b_2\By} \\
 \end{array}\right) }} \,,
\\
 &{\footnotesize {\arraycolsep=0.5mm
 \left(\begin{array}{cc} \BG^{(5)}_1 & \BG^{(5)}_2\\ \BG^{(5)}_3 & \BG^{(5)}_4 \end{array}\right)
 \equiv \left(\begin{array}{cc}
 m_{\alpha\beta} \Exp{3\Bphi}\BG^{a_1\cdots a_5,\,b_1\cdots b_5}
 & \sqrt{5}\Exp{3\Bphi}m_{\alpha\beta}\,\BG^{a_1\cdots a_5,\,b_1\cdots b_4\By} \\
 \sqrt{5}\Exp{3\Bphi}m_{\alpha\beta}\,\BG^{a_1\cdots a_4\By,\,b_1\cdots b_5}
 & 5\Exp{3\Bphi}m_{\alpha\beta}\,\BG^{a_1\cdots a_4\By,\,b_1\cdots b_4\By} \end{array}\right) }} \,,
\\
 &{\footnotesize {\arraycolsep=0.5mm
 \left(\begin{array}{cc} \BG^{(6,1)}_1 & \BG^{(6,1)}_2 \\ \BG^{(6,1)}_3 & \BG^{(6,1)}_4 \end{array}\right)
 \equiv 6\Exp{4\Bphi} \BG^{a_1\cdots a_6\By,\,b_1\cdots b_6\By}
 \left(\begin{array}{cc} \BG^{ab} & \BG^{a\By} \\ \BG^{\By b}
 & \BG^{\By\By} \end{array}\right) }} \,,
\\
\ell_2 = &{\footnotesize
 {\arraycolsep=0.7mm \left(\begin{array}{cc|cc|cc|cc|cc}
 0 & 0 & 0 & 0 & 0 & 0 & 0 & 0 & 0 & 0 \\
 0 & 0 & 0 & 0 & 0 & 0 & 0 & 0 & 0 & 0 \\ \hline
 \BB_{ab}^{\alpha} & \BB_{a\By}^{\alpha} & 0 & 0 & 0 & 0 & 0 & 0 & 
0 & 0 \\
 \BB_{\By b}^{\alpha} & 0 & 0 & 0 & 0 & 0 & 0 & 0 & 0 & 0 \\ \hline
 0 & 0
 &\frac{3!\,\epsilon_{\beta\gamma}\,
\delta_{a_1a_2a_3}^{bc_1c_2}\,\BB_{c_1c_2}^\gamma}{2!\sqrt{3!}}
 & 0 & 0 & 0 & 0 & 0 & 0 & 0 \\
 0 & 0
 &\frac{2!\,\epsilon_{\beta\gamma} \,\delta_{a_1a_2}^{bc}\,
 \BB_{c\By}^\gamma}{\sqrt{2!}}
 &\frac{\epsilon_{\beta\gamma}\,
 \BB_{a_1a_2}^\gamma}{\sqrt{2!}} & 0 & 0 & 0 & 0 & 0 & 0 \\ 
\hline
 0 & 0 & 0 & 0
 & \frac{5!\,\delta_{a_1\cdots a_5}^{b_1b_2b_3c_1c_2}\, \BB_{c_1c_2}^\alpha}{2!\sqrt{3!\,5!}} & 0 & 0 & 0 & 0 & 0 \\
 0 & 0 & 0 & 0 & \frac{4!\,\delta_{a_1\cdots a_4}^{b_1b_2b_3c} \,\BB_{c\By}^\alpha}{\sqrt{3!\,4!}} & \frac{4!\,\delta_{a_1\cdots a_4}^{b_1b_2c_1c_2}\, \BB_{c_1c_2}^\alpha}{2!\sqrt{2!\,4!}} & 0 & 0 & 0 & 0 \\ \hline
 0 & 0 & 0 & 0 & 0 & 0 & \, l^{(2)}_1 & l^{(2)}_2& 0 & 0 \\
 0 & 0 & 0 & 0 & 0 & 0 & \, l^{(2)}_3 & l^{(2)}_4 & \,0\, & \,0\,
 \end{array}\right)}} ,\\
 &{\footnotesize {\arraycolsep=0.5mm
 \left(\begin{array}{cc} l^{(2)}_1 & l^{(2)}_2\\ 
l^{(2)}_3&l^{(2)}_4\end{array}\right)
 \equiv \left(\begin{array}{cc}
 \epsilon^\rmT_{\beta\gamma} \delta^{b_1\cdots b_5}_{a_1\cdots a_5} \BB_{\By a}^\gamma
 & \frac{5!\,\epsilon^\rmT_{\beta\gamma}\, \delta^{b_1\cdots b_4c}_{a_1\cdots a_5} \BB_{a c}^\gamma}{\sqrt{4!\,5!}} \\
 0 & -\frac{5!\,\epsilon^\rmT_{\beta\gamma}\, \delta^{b_1\cdots b_4c}_{a_1\cdots a_5} \BB_{c\By}^\gamma}{\sqrt{4!\,5!}} \end{array}\right) }} \,,
\\[5pt]
\ell_4 = &{\footnotesize
 {\arraycolsep=0.5mm 
\left(\begin{array}{cc|cc|cc|@{~}c@{~}c@{~}|@{~}c@{~}c@{}}
 0 & 0 & 0 & 0 & 0 & 0 & 0 & 0 & 0 & 0 \\
 0 & 0 & 0 & 0 & 0 & 0 & 0 & 0 & 0 & 0 \\ \hline
 0 & 0 & 0 & 0 & 0 & 0 & 0 & 0 & 0 & 0 \\
 0 & 0 & 0 & 0 & 0 & 0 & 0 & 0 & 0 & 0 \\ \hline
 \frac{\BD_{a_1a_2a_3 b}}{\sqrt{3!}}
 & \frac{\BD_{a_1a_2a_3\By}}{\sqrt{3!}}
 & 0 & 0 & 0 & 0 & 0 & 0 & 0 & 0 \\
 \frac{\BD_{a_1a_2\By b}}{\sqrt{2!}}
 & 0 & 0 & 0 & 0 & 0 & 0 & 0 & 0 & 0 \\ \hline
 0 & 0
 &-\frac{5!\,\delta^\alpha_\beta\,
 \delta_{a_1\cdots a_5}^{bc_1\cdots c_4}\,
 \BD_{c_1\cdots c_4}}{4!\sqrt{5!}}
 & 0 & 0 & 0 & 0 & 0 & 0 & 0 \\
 0 & 0
 &-\frac{4!,\delta^\alpha_\beta\,
 \delta_{a_1\cdots a_4}^{bc_1c_2c_3}\,
 \BD_{c_1c_2c_3 \By}}{3!\sqrt{4!}}
 & -\frac{\delta^\alpha_\beta \, \BD_{a_1\cdots a_5}}{\sqrt{4!}}
 & 0 & 0 & 0 & 0 & 0 & 0 \\ \hline
 0 & 0 & 0 & 0 &l^{(4)}_1 & l^{(4)}_2& 0 & 0 & 0 & 0 \\
 0 & 0 & 0 & 0 & l^{(4)}_3&l^{(4)}_4 & 0 &0 & 0 & 0
 \end{array}\right) }},
\\
 &{\footnotesize {\arraycolsep=0.5mm
 \left(\begin{array}{cc} l^{(4)}_1 & l^{(4)}_2\\
 l^{(4)}_3&l^{(4)}_4\end{array}\right)
 \equiv
 \left(\begin{array}{cc}
 -\frac{5!\,\delta^{b_1b_2b_3c_1c_2}_{a_1\cdots a_5}\,
 \BD_{c_1c_2\By \alpha}}{2!\sqrt{3!\,5!}}
 &-\frac{5!\,\delta^{b_1b_2c_1c_2c_3}_{a_1\cdots a_5}\,
 \BD_{ac_1c_2c_3}}{3!\sqrt{2!\,5!}} \\
 0 & \frac{5!\,\delta^{b_1b_2c_1c_2c_3}_{a_1\cdots a_5}\,
 \BD_{c_1c_2c_3\By}}{3!\sqrt{2!\,5!}}
 \end{array}\right) }} \,,
\\[5pt]
\ell_6= &{\footnotesize
 {\arraycolsep=0.5mm \left(\begin{array}{cc|cc|cc|cc|cc}
 0 & 0 & 0 & 0 & 0 & 0 & 0 & 0 & 0 & 0 \\
 0 & 0 & 0 & 0 & 0 & 0 & 0 & 0 & 0 & 0 \\ \hline
 0 & 0 & 0 & 0 & 0 & 0 & 0 & 0 & 0 & 0 \\
 0 & 0 & 0 & 0 & 0 & 0 & 0 & 0 & 0 & 0 \\ \hline
 0 & 0 & 0 & 0 & 0 & 0 & 0 & 0 & 0 & 0 \\
 0 & 0 & 0 & 0 & 0 & 0 & 0 & 0 & 0 & 0 \\ \hline
 0 & \frac{\BD_{a_1\cdots a_5 \By}^\alpha}{\sqrt{5!}} & 0 & 0 & 0 & 
0 & 0 & 0 & 0 & 0 \\
 -\frac{\BD_{a_1\cdots a_4 b \By}^\alpha}{\sqrt{4!}} & 0 & 0 & 0 & 0 
& 0 & 0 & 0 & 0 & 0 \\ \hline
 0 & 0 & \frac{5!\,\epsilon^\rmT_{\beta\gamma}\,\delta^{bc_1\cdots 
c_4}_{a_1\cdots a_5}\BD_{a c_1\cdots c_4\By}^\gamma}{4!\sqrt{5!}} & 0 
& 0 & 0 & 0 & 0 & 0 & 0 \\
 0 & 0 & 0 & \frac{\epsilon^\rmT_{\beta\gamma}\,\BD_{a_1\cdots 
a_5\By}^\gamma}{\sqrt{5!}} & \,0\, & \,0\, & \,0\, & \,0\, & \,0\, & 
\,0\,
 \end{array}\right)}} \,.
\end{align}

\subsection{Similarity transformation}

In order to examine the map \eqref{eq:linear-map}, it is convenient to use \eqref{eq:M-IIB-map} from the beginning, although we have found \eqref{eq:M-IIB-map}.
Using \eqref{eq:M-IIB-map}, the metric and the inverse metric become
\begin{align}
 \bigl(G_{ij}\bigr)
 &= \Exp{-\frac{2}{3}\,\Bphi}\,\BG_{\By\By}^{1/3}
 \begin{pmatrix}
 \delta_a^c & -\BB_{a\By}^\gamma \\ 0 & \delta_\alpha^\gamma
 \end{pmatrix}
\begin{pmatrix}
 \frac{2\,\BG_{c\By,\,d\By}}{\BG_{\By\By}} & 0 \\ 0 & 
\frac{\Exp{\Bphi}}{\BG_{\By\By}}\, m_{\gamma\delta}
 \end{pmatrix}
 \begin{pmatrix}
 \delta^d_b & 0 \\ -\BB_{b\By}^\delta & \delta^\delta_\beta
 \end{pmatrix} \,,
\\
 \bigl(G^{ij}\bigr) &=
 \Exp{\frac{2}{3}\,\Bphi}\,\BG_{\By\By}^{-1/3}
 \begin{pmatrix}
 \delta_a^c & 0 \\ \BB_{a\By}^\gamma & \delta_\alpha^\gamma
 \end{pmatrix}
\begin{pmatrix}
 \BG^{cd} & 0 \\ 0 & \Exp{-\Bphi}\BG_{\By\By}\, m^{\gamma\delta}
 \end{pmatrix}
 \begin{pmatrix}
 \delta^d_b & \BB_{b\By}^\delta \\ 0 & \delta^\delta_\beta
 \end{pmatrix}\,.
\end{align}
If we define
\begin{align}
 \bigl(\overline{G}_{ij}\bigr)&\equiv 
\Exp{-\frac{2}{3}\,\Bphi}\,\BG_{\By\By}^{1/3}
 \begin{pmatrix}
 \frac{2\,\BG_{a\By,\,b\By}}{\BG_{\By\By}} & 0 \\ 0 & 
\frac{\Exp{\Bphi}}{\BG_{\By\By}}\, m_{\alpha\beta}
 \end{pmatrix} \,, \quad
 (\Lambda^i{}_j)\equiv \begin{pmatrix}
 \delta^a_b & 0 \\ -\BB_{b\By}^\alpha & \delta^\alpha_\beta
 \end{pmatrix} \,,
\end{align}
they satisfy $G_{ij} = \Lambda^k{}_i\,\Lambda^l{}_j\,\overline{G}_{kl}$\,.
Then, we have $\hat{\cM}_{IJ}=(\Sigma^\rmT)_I{}^K\,\overline{\cM}_{KL}\,\Sigma^L{}_J$ with
\begin{align}
 (\overline{\cM}_{IJ}) &\equiv \abs{G}^{\frac{1}{n-2}}
 \begin{pmatrix}
 \overline{G}_{ij} & 0 & 0 & 0 \\
 0 & \overline{G}^{i_1i_2,\,j_1j_2} & 0 & 0 \\
 0 & 0 & \overline{G}^{i_1\cdots i_5,\,j_1\cdots j_5} & 0 \\
 0 & 0 & 0 & \overline{G}^{i_1\cdots i_7,\,j_1\cdots 
j_7}\,\overline{G}^{ij}
 \end{pmatrix}\,,
\\
 \bigl(\Sigma^I{}_J\bigr)&\equiv
 \begin{pmatrix}
 \Lambda^i{}_j & 0 & 0 & 0 \\
 0 & \bigl(\Lambda^{-\rmT}\bigr)_{i_1i_2}^{\,\,\,j_1j_2} & 0 & 0 \\
 0 & 0 & \bigl(\Lambda^{-\rmT}\bigr)_{i_1\cdots 
i_5}^{\,\,\,j_1\cdots j_5} & 0 \\
 0 & 0 & 0 & \bigl(\Lambda^{-\rmT}\bigr)_{i_1\cdots 
i_7}^{\,\,\,j_1\cdots j_7}\,\bigl(\Lambda^{-\rmT}\bigr)_i{}^j
 \end{pmatrix}\,,
\end{align}
where $\bigl(\Lambda^{-\rmT}\bigr)_{i_1\cdots i_p}^{\,\,\,j_1\cdots j_p} \equiv \bigl(\Lambda^{-1}\bigr)^{[j_1}{}_{[i_1}\cdots \bigl(\Lambda^{-1}\bigr)^{j_p]}{}_{i_p]}$\,.
In terms of the type IIB fields, the non-vanishing components of $\overline{G}^{i_1\cdots i_p,\,j_1\cdots j_p}$ and $\bigl(\Lambda^{-\rmT}\bigr)_{i_1\cdots i_p}^{\,\,\,j_1\cdots j_p}$, and also $\abs{G}$ are given by
\begin{align}
\begin{split}
 &\overline{G}^{a_1\cdots a_p,\,b_1\cdots b_p} = 
\Exp{\frac{2p}{3}\,\Bphi}\,\BG_{\By\By}^{-p/3}\,\BG^{a_1\cdots 
a_p,\,b_1\cdots b_p}\,,
\\
 &\overline{G}^{a_1\cdots a_{p-1}\alpha,\,b_1\cdots b_{p-1}\beta} = 
\frac{1}{p}\,\Exp{\frac{2p-3}{3}\,\Bphi}\,\BG_{\By\By}^{(3-p)/3}\,\BG^{a_1\cdots 
a_{p-1},\,b_1\cdots b_{p-1}}\, m^{\alpha\beta} \,,
\\
 &\overline{G}^{a_1\cdots a_{p-2}\Ay\Az,\,b_1\cdots b_{p-2}\Ay\Az} = 
\frac{1}{p(p-1)}\,\Exp{\frac{2p-6}{3}\,\Bphi}\,\BG_{\By\By}^{(6-p)/3}\,\BG^{a_1\cdots 
a_{p-2},\,b_1\cdots b_{p-2}}\,\,,
\\
 &\bigl(\Lambda^{-\rmT}\bigr)_{a_1\cdots a_p}^{\,\,\,b_1\cdots b_p} 
= \delta_{a_1\cdots a_p}^{b_1\cdots b_p}\,, \quad
 \bigl(\Lambda^{-\rmT}\bigr)_{a_1\cdots a_p}^{\,\,\,b_1\cdots 
b_{p-1}\beta} = \delta_{[a_1\cdots a_{p-1}}^{b_1\cdots b_{p-1}}\,\BB_{a_p]\By}^\beta\,,
\\
 &\bigl(\Lambda^{-\rmT}\bigr)_{a_1\cdots a_{p-1}\alpha}^{\,\,\,b_1\cdots b_{p-1}\beta} 
 = \frac{1}{p}\,\delta_{a_1\cdots a_{p-1}}^{b_1\cdots b_{p-1}}\,\delta_\alpha^\beta\,, \quad
 \bigl(\Lambda^{-\rmT}\bigr)_{a_1\cdots a_p}^{\,\,\,b_1\cdots b_{p-2}\Ay\Az} 
 = \frac{1}{2}\,\epsilon_{\gamma\delta}\,\delta_{[a_1\cdots a_{p-2}}^{b_1\cdots b_{p-2}}\,\BB_{a_{p-1}|\By|}^\gamma\,\BB_{a_p]\By}^\delta\,,
\\
 &\bigl(\Lambda^{-\rmT}\bigr)_{a_1\cdots a_{p-1}\alpha}^{\,\,\,b_1\cdots b_{p-2}\Ay\Az} 
 = \frac{1}{p}\,\epsilon^\rmT_{\alpha\gamma}\,\delta_{[a_1\cdots a_{p-2}}^{b_1\cdots b_{p-2}}\,\BB^\gamma_{a_{p-1}]\By}\,, \quad
 \bigl(\Lambda^{-\rmT}\bigr)_{a_1\cdots a_{p-2}\Ay\Az}^{\,\,\,b_1\cdots b_{p-2}\Ay\Az} 
 = \frac{1}{p(p-1)}\,\delta_{a_1\cdots a_{p-2}}^{b_1\cdots b_{p-2}} \,.
\end{split}
\end{align}
We can also show that
\begin{align}
\begin{split}
 &\abs{G}^{\frac{1}{n-2}} = \abs{\overline{G}}^{\frac{1}{n-2}} = 
\Exp{\frac{2}{3}\,\Bphi}\BG_{\By\By}^{-\frac{1}{3}} 
\Exp{-\frac{4}{n-2}\,\Bphi} \abs{\BG}^{\frac{1}{n-2}}\,,
\\
 &\abs{\BG}\equiv \det(\BG_{\sfm\sfn}) \,, \quad
 \bigl(\BG_{\sfm\sfn}\bigr)\equiv \begin{pmatrix}
 \BG_{ab} & \BG_{a\By} \\ \BG_{b\By} & \BG_{\By\By}
 \end{pmatrix} \,.
\end{split}
\end{align}

Under the $(d-2)+2$ decomposition, we obtain
\begin{align}
\overline{\cM} =&\Exp{-\frac{4}{n-2}\,\Bphi} \abs{\BG}^{\frac{1}{n-2}}
 {\footnotesize
 {\arraycolsep=2pt 
\left(\begin{array}{@{}c@{}c|c@{}c@{}c|c@{}c@{}c|c@{}c@{}}
 \frac{2\BG_{a\By, b\By}}{\BG_{\By\By}} & 0 & 0 & 0 & 0 & 0 & 0 & 0 
& 0 & 0 \\
 0 & \frac{\Exp{\Bphi}m_{\alpha\beta}}{\BG_{\By\By}}\, & 0 & 0 & 0 
& 0 & 0 & 0 & 0 & 0 \\ \hline
 0 & 0 & \,\frac{\Exp{2\Bphi} \BG^{a_1a_2,b_1b_2}}{\BG_{\By\By}} & 
0 & 0 & 0 & 0 & 0 & 0 & 0 \\
 0 & 0 & 0 & \Exp{\Bphi}\BG^{ab} m^{\alpha\beta} & 0 & 0 & 0 & 0 & 
0 & 0 \\
 0 & 0 & 0 & 0 & \BG_{\By\By}\, & 0 & 0 & 0 & 0 & 0 \\ \hline
 0 & 0 & 0 & 0 & 0 & \,\overline{m}^{(5)}_1 & 0 & 0 & 0 & 0 \\
 0 & 0 & 0 & 0 & 0 & 0 &\overline{m}^{(5)}_2& 0 & 0 & 0 \\
 0 & 0 & 0 & 0 & 0 & 0 & 0 & \overline{m}^{(5)}_3& 0 & 0 \\ \hline
 0 & 0 & 0 & 0 & 0 & 0 & 0 & 0 & \,\overline{m}^{(7,1)}_1& 0 \\
 0 & 0 & 0 & 0 & 0 & 0 & 0 & 0 & 0 &\overline{m}^{(7,1)}_2
 \end{array}\right) }\,, }
\\
&{\arraycolsep=0mm
 \left(\begin{array}{ccc}
 \overline{m}^{(5)}_1&0&0\\
 0& \overline{m}^{(5)}_2&0\\
 0&0& \overline{m}^{(5)}_3
 \end{array}\right)}
 \equiv \left(\begin{array}{@{}*3{c@{}}}
 \frac{\Exp{4\Bphi}\BG^{a_1\cdots a_5,b_1\cdots 
b_5}}{\BG_{\By\By}^2}
 &0&0\\
 0&\frac{\Exp{3\Bphi}m^{\alpha\beta}
 \BG^{a_1\cdots a_4,b_1\cdots b_4}}{\BG_{\By\By}}&0\\
 0&0&\Exp{2\Bphi} \BG^{a_1a_2a_3,b_1b_2b_3}\\
\end{array}\right),
\\
&{\arraycolsep=0mm
 \left(\begin{array}{cc}
 \overline{m}^{(7,1)}&0\\ 0& \overline{m}^{(7,1)}_2
 \end{array}\right)}
 \equiv{\arraycolsep=0mm\left(\begin{array}{cc}
 \frac{\Exp{4\Bphi}\BG^{a_1\cdots a_5,b_1\cdots b_5}
 \BG^{ab}}{\BG_{\By\By}}&0 \\
 0&\Exp{3\Bphi}m^{\alpha\beta}\BG^{a_1\cdots a_5,b_1\cdots b_5}
 \end{array}\right)}\,,
\\
 \Sigma = &{\footnotesize
 {\arraycolsep=1mm \left(\begin{array}{@{}cc|ccc|ccc|cc@{}}
 \delta^a_b & 0 & 0 & 0 & 0 & 0 & 0 & 0 & 0 & 0 \\
 -\BB^\alpha_{b\By} & \delta^\alpha_\beta\,
 & 0 & 0 & 0 & 0 & 0 & 0 & 0 & 0 \\ \hline
 0 & 0 & \,\delta_{a_1a_2}^{b_1b_2}
 &\frac{2!\,\delta_{a_1a_2}^{bc}\,\BB^\beta_{c\By}}{\sqrt{2!}}
 &\frac{2!\,\epsilon_{\gamma\delta}\,\delta_{a_1a_2}^{c_1c_2}\, \BB_{c_1\By}^\gamma\,\BB_{c_2 \By}^\delta}{2!\sqrt{2!}}\,
 & 0 & 0 & 0 & 0 & 0 \\
 0 & 0 & 0 & \delta_a^b\,\delta_\alpha^\beta & \epsilon^\rmT_{\alpha\gamma}\,\BB_{a\By}^\gamma
 & 0 & 0 & 0 & 0 & 0 \\
 0 & 0 & 0 & 0 & 1 & 0 & 0 & 0 & 0 & 0 \\ \hline
 0 & 0 & 0 & 0 & 0 &s^{(5)}_1&s^{(5)}_2&s^{(5)}_3 & 0 & 0 \\
 0 & 0 & 0 & 0 & 0 & 0 &s^{(5)}_4&s^{(5)}_5& 0 & 0 \\
 0 & 0 & 0 & 0 & 0 & 0 & 0 &s^{(5)}_6& 0 & 0 \\ \hline
 0 & 0 & 0 & 0 & 0 & 0 & 0 & 0 &s^{(7,1)}_1&s^{(7,1)}_2\\
 0 & 0 & 0 & 0 & 0 & 0 & 0 & 0 &0& s^{(7,1)}_3
 \end{array}\right)}},
\\
 & \begin{pmatrix}
 s^{(5)}_1 & s^{(5)}_2 & s^{(5)}_3 \\
 0 & s^{(5)}_4 & s^{(5)}_5 \\
 0 & 0 & s^{(5)}_6 \end{pmatrix}
 \equiv \begin{pmatrix}
 \delta_{a_1\cdots a_5}^{b_1\cdots b_5} & \frac{5!\,\delta_{a_1\cdots a_5}^{b_1\cdots b_4c}\, \BB^\beta_{c\By}}{\sqrt{4!\,5!}} & \frac{5!\,\epsilon_{\gamma\delta}\, \delta_{a_1\cdots a_5}^{b_1b_2b_3c_1c_2}\, \BB^\gamma_{c_1\By}\,\BB_{c_2\By}^\delta}{2!\sqrt{3!\,5!}} \\
 0 & \delta_\alpha^\beta\,\delta_{a_1\cdots a_4}^{b_1\cdots b_4} & \frac{4!\,\epsilon^\rmT_{\alpha\gamma}\, \delta^{b_1b_2b_3c}_{a_1\cdots a_4}\,\BB^\gamma_{c\By}}{\sqrt{3!\,4!}} \\
 0&0&\delta_{a_1a_2a_3}^{b_1b_2b_3}
\end{pmatrix} ,
\\
 &\begin{pmatrix} s^{(7,1)}_1&s^{(7,1)}_2\\ 0& s^{(7,1)}_3 \end{pmatrix}
 \equiv \begin{pmatrix} \delta_{a_1\cdots a_5}^{b_1\cdots b_5} \delta_a^b & \delta_{a_1\cdots a_5}^{b_1\cdots b_5}\,\BB_{a\By}^\beta \\
 0 & \delta_{a_1\cdots a_5}^{b_1\cdots b_5}\,\delta_\alpha^\beta
 \end{pmatrix} \,.
\end{align}
$L_3$ and $L_6$ can also be decomposed in a similar manner. 
From these, we can perform the similarity transformation,
\begin{align}
 \cM= L_6^\rmT\,L_3^\rmT\,\Sigma^\rmT\,\overline{\cM} \,\Sigma\, 
L_3\,L_6
 \quad\to\quad S^{\rmT} \cM \,S \,.
\end{align}
After laborious calculations, one can show that $S^{\rmT} \cM \,S$ is 
equal to \eqref{eq:IIB-decomposed}.

\end{document}